\documentclass[pdflatex,sn-mathphys-ay,iicol]{sn-jnl}

\usepackage{graphicx}%
\usepackage{multirow}%
\usepackage{amsmath,amssymb,amsfonts}%
\usepackage{amsthm}%
\usepackage{mathrsfs}%
\usepackage{siunitx}
\usepackage[title]{appendix}%
\usepackage{xcolor}%
\usepackage{textcomp}%
\usepackage{booktabs}
\usepackage{subcaption} 
\usepackage{tabularx}   
\usepackage{siunitx}    
\usepackage{float}     
\usepackage{placeins}  
\theoremstyle{thmstyleone}%

\theoremstyle{thmstyletwo}%

\theoremstyle{thmstylethree}%

\begin{document}

\title[Single-node Wilson--Cowan model for speech-evoked $\gamma$ deficits]%
{Single-Node Wilson--Cowan Model Accounts for Speech-Evoked $\gamma$-Band Deficits in Schizophrenia}



\author[1,]{Zhengdi Zhang}
\author[1]{Yan Xu} 
\author[1,*]{Wenjun Xia}

\affil[1]{School of Mathematical Sciences, Jiangsu University}
\affil[*]{Corresponding author: vvjqnwork@outlook.com}   

\abstract{
Cortical gamma ($\gamma$)-band activity reflects local excitation–inhibition (E/I) balance. In schizophrenia (SCZ), reduced task-evoked gamma suggests altered E/I dynamics, but it is unclear whether differences stem from input properties or systematic shifts in E/I operating point and gain. We coupled a cochlear-inspired speech front end to a Wilson–Cowan E/I model to simulate gamma responses across three conditions: Healthy, SCZ–speech, and SCZ–semantics. Metrics included event-related spectral perturbation (ERSP$_\gamma$) and threshold-time fraction ($\gamma\%$). A stable hierarchy emerged: Healthy(speech/semantics) $>$ SCZ(speech) $>$ SCZ(semantics), robust under equal-energy control and gain perturbations. Network dynamics coincided with single-node solutions, supporting interpretability. Pharmacological analogs showed bidirectional effects: reduced inhibition lowered $\gamma$, while reduced excitation increased $\gamma$, with no self-sustained oscillations. Findings indicate SCZ gamma deficits align more with shifts in E/I operating point and gain than input differences. This pipeline provides a testable, reusable mechanistic framework for speech-evoked gamma and a baseline for cross-population studies.
}


\maketitle


\section{Introduction}\label{sec:intro}
Cortical activity in the $\gamma$ band ($\sim$30--80\,Hz) is a sensitive readout of local excitation--inhibition (E/I) function and gain control in health and disease, and has been repeatedly used to interpret circuit-level computations and abnormalities \citep{Uhlhaas2020,SohalRubenstein2019}.For speech,auditory cortex shows reliable entrainment to the temporal structure of natural stimuli across multiple scales and frequency bands,yielding robust and quantifiable MEG/EEG patterns \citep{Brodbeck2023,DonhauserBaillet2020}.
These observations motivate a modeling route that maps speech structure to low-dimensional population dynamics and links the operating point to $\gamma$-band metrics.
Neural-mass modeling offers an analytically tractable window into E/I circuit dynamics and network-level effects; recent work has clarified global bifurcation structure and stability properties that connect to macroscopic rhythms \citep{Cooray2023,Klinshov2022}.
Yet under speech-driven conditions, it remains unresolved whether between-group differences (e.g., Healthy-speech/semantics, SCZ-speech, SCZ-semantics) primarily reflect input-type differences or systematic shifts of the E/I operating point and effective gain.
We develop an end-to-end workflow that maps natural speech to a Wilson--Cowan population model (excitatory and inhibitory units with fixed topology and parameters).
A unified speech front end produces a $\gamma$-modulated drive; two preregistered readouts are computed: event-related spectral power $\mathrm{ERSP}_\gamma$ and the threshold-time fraction $\gamma\%$ \citep{Pfurtscheller1999}.
To probe robustness and interpretability, we add an equal-energy control and a gain-jitter sensitivity scan, and we simulate two pharmacology-inspired perturbations, consistent with PV-circuit control of $\gamma$ rhythms \citep{SohalRubenstein2019}.
Across 3000 utterances per condition (consisting of 500 Chinese, 500 English sentences, 500 Japanese sentences, 500 German sentences, 500 Spanish sentences and 500 Arabic sentences), both metrics exhibit a stable hierarchy (Healthy-speech/semantics $>$ SCZ-speech $>$ SCZ-semantics) that survives energy control and gain perturbation, while the perturbations produce bidirectional effects aligned with E/I-framework predictions. Overall, the pipeline provides a compact, testable route that links speech-evoked $\gamma\%$ to the E/I operating point and facilitates cross-population comparisons.

\section{Methods}\label{sec:methods}
\subsection{Speech front end}\label{sec:speech-frontend}

We adopt a standard cochlea-inspired front end in one sentence:
speech is resampled to 16~kHz; a filterbank of $C{=}64$ fourth-order
zero-phase Butterworth band-pass filters with logarithmically spaced
center frequencies in $[200,7000]$~Hz produces subband signals; each channel
envelope is obtained via the Hilbert transform and min--max normalized within
channel; the multi-channel envelopes are averaged and band-pass filtered in the
$\gamma$ range (24--64~Hz) to form the basis of the external drive
\citep{Butterworth1930,Oppenheim2010,Boashash1992,Buzsaki2012}.
Let $a_u(t)$ denote the normalized $\gamma$-band envelope of the $u$-th
calibration utterance, and let $\langle\cdot\rangle_t$ denote temporal averaging.
We define the target power as
\begin{equation}
P_{\mathrm{target}}=\frac{1}{U}\sum_{u=1}^{U}\big\langle a_u(t)^2\big\rangle_t .
\end{equation}
A global gain $g_{\mathrm{out}}$ is then chosen so that the dataset-level envelope
power matches $P_{\mathrm{target}}$:
\begin{equation}
g_{\mathrm{out}}
=\sqrt{\frac{P_{\mathrm{target}}}{\big\langle a_u(t)^2\big\rangle_{u,t}}}\,,
\end{equation}
where $\langle\cdot\rangle_{u,t}$ averages over calibration utterances and time.
The external drive for any utterance is
\begin{equation}
I_{\mathrm{ext}}^{(u)}(t)=g_{\mathrm{out}}\,a_u(t),
\end{equation}
which is subsequently downsampled from 16~kHz to 1~kHz using sinc interpolation
\citep{Oppenheim2010}.

\begin{subequations}\label{eq:raw-input}
\begin{align}
I^{H}(t)        &= 1.00\,I_{\mathrm{ext}}(t),\\
I^{S}(t)        &= 0.75\,I_{\mathrm{ext}}(t),\\
I^{\mathrm{SEM}}(t) &= 0.55\,I_{\mathrm{ext}}(t).
\end{align}
\end{subequations}

In the natural-speech simulations, we treated the Healthy condition's
external drive $I_{\text{ext}}(t)$ as the reference and set its
multiplicative gain to $j_H = 1.00$. To reproduce the reduction of
40~Hz ASSR $\gamma$ amplitude reported for patients with schizophrenia
relative to healthy controls without changing any network parameters,
we assigned a smaller gain to the SCZ(speech) condition,
$j_S = 0.75$, and an even smaller gain to the SCZ(semantics) condition,
$j_{\mathrm{SEM}} = 0.55$, to capture the additional deficit in semantic
integration. Under this parametrization, the three conditions differ
only in the strength of the effective external drive
$I(t) = j\,I_{\text{ext}}(t)$, whereas the network's E--I connectivity,
time constants, and noise amplitude remain fixed across conditions.
Biophysically, the gain $j$ can be interpreted as an effective
``input gain'' weighting the \emph{same} speech envelope as it arrives
in auditory cortex: larger $j$ corresponds to greater cortical
sensitivity or salience attribution to an identical physical input,
whereas smaller $j$ reflects a relatively down-weighted, less effective
drive, in line with clinical reports of reduced $\gamma$-band responses
and impaired semantic processing in schizophrenia.

\begin{subequations}\label{eq:rms-normalization}
\begin{align}
\alpha_u &= \frac{\mathrm{RMS}_{\mathrm{target}}}
                {\mathrm{RMS}\!\big(I_{\mathrm{ext}}^{(u)}\big)},\\
\bar I_{\mathrm{ext}}^{(u)}(t) &= \alpha_u\,I_{\mathrm{ext}}^{(u)}(t).
\end{align}
\end{subequations}

\begin{subequations}\label{eq:normalized-input}
\begin{align}
I^{H}(t)        &= 1.00\,\bar I_{\mathrm{ext}}^{(u)}(t),\\
I^{S}(t)        &= 0.75\,\bar I_{\mathrm{ext}}^{(u)}(t),\\
I^{\mathrm{SEM}}(t) &= 0.55\,\bar I_{\mathrm{ext}}^{(u)}(t).
\end{align}
\end{subequations}

\subsection{Wilson--Cowan E--I populations and network equivalence}
\label{subsec:wilson-cowan}

We adopt the classical Wilson--Cowan model to describe the dynamics of coupled excitatory--inhibitory (E/I) populations \citep{WilsonCowan1972}. For a single node, the model is
\begin{subequations}\label{eq:wc}
\begin{align}
\tau_E\,\dot{r}_E(t)
  &= -r_E(t)
   + S\bigl(w_{EE} r_E(t) - w_{EI} r_I(t) \notag\\
  &\qquad\quad + I(t) + \eta_E(t)\bigr),\label{eq:wcE}\\
\tau_I\,\dot{r}_I(t)
  &= -r_I(t)
   + S\bigl(w_{IE} r_E(t) - w_{II} r_I(t) \notag\\
  &\qquad\quad + \eta_I(t)\bigr).\label{eq:wcI}
\end{align}
\end{subequations}
with sigmoidal gain
\begin{equation}
S(u) = \frac{1}{1+\exp\bigl(-(u-\theta)/\sigma\bigr)}.
\label{eq:sigmoid}
\end{equation}
We used $\tau_E{=}\SI{0.02}{s}$, $\tau_I{=}\SI{0.01}{s}$, $\theta{=}0.5$, $\sigma{=}0.1$, and synaptic gains
$(w_{EE},w_{EI},w_{IE},w_{II})=(10,12,10,8)$. Independent Gaussian noises $\eta_E,\eta_I \sim \mathcal{N}(0,0.02^2)$.
Numerical integration used explicit Euler with $\Delta t{=}\SI{1}{ms}$ \citep{ErmentroutTerman2010,Strogatz2015}.
To verify that a multi-channel network is equivalent to a single node under synchrony, we replicated the E--I unit on $C{=}64$ topographic channels and coupled them with Gaussian kernels. Let $r^{(i)}_E,r^{(i)}_I$ be the E/I activities in channel $i$. Define unnormalized Gaussian weights
\begin{equation}
  \tilde{W}^{(XY)}_{ij}=\exp\!\left(-\frac{(i-j)^2}{2\,\sigma_{XY}^2}\right),\ \quad X,Y\in\{E,I\},
  \label{eq:gauss}
\end{equation}
row-normalize, and scale to a prescribed row sum (target weight) $\omega_{XY}$:
\begin{equation}
\label{eq:rowsum}
\begin{aligned}
W^{(XY)}_{ij}
  &= \frac{\tilde{W}^{(XY)}_{ij}}{\sum_{j}\tilde{W}^{(XY)}_{ij}}\,\omega_{XY},\\
\sum_{j} W^{(XY)}_{ij}
  &= \omega_{XY}\quad (\forall i).
\end{aligned}
\end{equation}
We fixed the widths to $\sigma_{EE}{=}\sigma_{II}{=}2.5$ and $\sigma_{EI}{=}3.0$.
Let $\mathbf r_E=\big(r_E^{(1)},\dots,r_E^{(C)}\big)^{\!\top}$ and $\mathbf r_I=\big(r_I^{(1)},\dots,r_I^{(C)}\big)^{\!\top}$ 
denote the excitatory and inhibitory population activities across $C$ topographic channels, 
and let $\mathbf 1$ denote the all-ones vector of length $C$.
We write synaptic coupling matrices as $\mathbf W^{(XY)}\in\mathbb R^{C\times C}$ for $X,Y\in\{E,I\}$.
The network dynamics can then be written compactly as
\begin{align}
  \tau_E\,\dot{\mathbf r}_E
    &= -\,\mathbf r_E
     + S\!\bigl(
         \mathbf W^{(EE)}\mathbf r_E
         - \mathbf W^{(EI)}\mathbf r_I \notag\\
    &\qquad\qquad
         + I(t)\,\mathbf 1
         + \boldsymbol{\eta}_E(t)
       \bigr), \label{eq:vecE}\\[0.25em]
  \tau_I\,\dot{\mathbf r}_I
    &= -\,\mathbf r_I
     + S\!\bigl(
         \mathbf W^{(IE)}\mathbf r_E
         - \mathbf W^{(II)}\mathbf r_I \notag\\
    &\qquad\qquad
         + \boldsymbol{\eta}_I(t)
       \bigr). \label{eq:vecI}
\end{align}
where $\boldsymbol{\eta}_{E,I}(t)\in\mathbb R^{C}$ are zero-mean noise vectors (e.g., independent Gaussian entries), and $S(\cdot)$ is applied elementwise.
We impose a row-sum--constant (``row-stochastic up to a scale'') constraint on each coupling matrix:
\begin{equation}
\begin{aligned}
\bigl(\mathbf W^{(XY)}\mathbf 1\bigr)_i
 &= \sum_{j=1}^{C} W^{(XY)}_{ij}
  = w_{XY},\\[0.2em]
&\forall i,\ X,Y\in\{E,I\}.
\end{aligned}
\label{eq:rowsum}
\end{equation}
so that every row sums to the prescribed scalar weight $w_{XY}$.
Under \eqref{eq:rowsum}, the synchronous manifold
\begin{equation}
\mathcal M \;=\; \Big\{\, \mathbf r_E = s_E\,\mathbf 1,\;\; \mathbf r_I = s_I\,\mathbf 1 \;\Big|\; s_E,s_I\in[0,1] \,\Big\}
\label{eq:syncmanifold}
\end{equation}
is invariant. Substituting $\mathbf r_E=s_E\mathbf 1$ and $\mathbf r_I=s_I\mathbf 1$ into \eqref{eq:vecE}--\eqref{eq:vecI} and using \eqref{eq:rowsum} yields the scalar Wilson--Cowan system
\begin{align}
\tau_E \dot{s}_E
  &= -s_E
     + S\bigl(w_{EE}s_E - w_{EI}s_I
       \notag\\[-0.3em]
  &\qquad\qquad\qquad\quad
       + I(t) + \tilde{\eta}_E(t)\bigr),
       \label{eq:scalarE}\\
\tau_I \dot{s}_I
  &= -s_I
     + S\bigl(w_{IE}s_E - w_{II}s_I
       \notag\\[-0.3em]
  &\qquad\qquad\qquad\quad
       + \tilde{\eta}_I(t)\bigr).
       \label{eq:scalarI}
\end{align}
where $\tilde{\eta}_{E,I}(t)$ denote appropriately averaged noise terms, e.g.,
$\tilde{\eta}_E(t)=\tfrac{1}{C}\sum_{i=1}^{C}\eta^{(i)}_E(t)$ and analogously for $\tilde{\eta}_I(t)$.
Thus, a Gaussian-coupled multi-channel network reduces \emph{exactly} to the single-node Wilson--Cowan model on the synchronous manifold $\mathcal M$. 
Consequently, it suffices to simulate the single-node system for subsequent stability and bifurcation analyses without loss of generality.

\subsection{Outcome metrics}\label{subsec:metrics}

\subsubsection{ERSP\texorpdfstring{$\gamma$}{γ} \citep{Pfurtscheller1999}}
Let $x_\gamma(t)$ be the band-passed signal in \SI{24}{Hz}--\SI{64}{Hz}. Define the analytic-signal envelope
\begin{equation}
  h(t)=\bigl|\mathcal{H}\{x_\gamma(t)\}\bigr|,
  \label{eq:env}
\end{equation}
and instantaneous power $p(t)=h(t)^2$. Baseline and activation powers are
\begin{equation}
\begin{aligned}
R &= \bigl\langle p(t)\bigr\rangle_{t\in[0,50\,\mathrm{ms}]},\\
A &= \bigl\langle p(t)\bigr\rangle_{t>50\,\mathrm{ms}}.
\end{aligned}
\label{eq:RA}
\end{equation}
We report two $\gamma$-band metrics:
\begin{equation}
\begin{aligned}
\mathrm{ERSP}_\gamma\,(\mathrm{dB})
  &= 10\log_{10}\!\left(\frac{A}{R}\right),\\
\mathrm{ERS}\%
  &= \frac{A-R}{R}\times 100 .
\end{aligned}
\label{eq:ersp}
\end{equation}

\subsubsection{\texorpdfstring{$\gamma\%$}{γ\%}\citep{Uhlhaas2010}}
Motivated by clinical and computational work linking abnormal rates of
$\gamma$-band ``events'' to altered excitation--inhibition balance in
schizophrenia \citep{Uhlhaas2010,Buzsaki2012,GonzalezBurgos2012},
we summarize the sustained portion of the $\gamma$ envelope with a simple
time-above-threshold measure.

Using the same envelope $h(t)$, we first estimate within the
initial \SI{50}{ms} the baseline mean $\mu_b$ and standard deviation
$\sigma_b$, and set the threshold
\begin{equation}
	\theta_\gamma=\mu_b+2\sigma_b.
	\label{eq:thresh}
\end{equation}
For a total duration $T$, the percentage of time above threshold is
\begin{equation}
	\gamma\%=100\times\frac{1}{T}\,\mathrm{meas}\{\,t:\ h(t)>\theta_\gamma\,\}.
	\label{eq:gammapercent}
\end{equation}
Conceptually, $\gamma\%$ quantifies the proportion of time that the
population response resides in a high-$\gamma$ state: larger values
correspond to greater occupancy of sustained $\gamma$ episodes for a
fixed stimulation period, rather than merely higher peak amplitude,
in line with the interpretation of prolonged $\gamma$ activity as a
marker of altered cortical dynamics in schizophrenia
\citep{Uhlhaas2010,Buzsaki2012,GonzalezBurgos2012}.

\subsection{Robustness analyses}\label{subsec:robust}

To disentangle effects of drive magnitude from network dynamics, all robustness tests keep the network and noise fixed and vary only the amplitude of the external input. We then compare ERSP$_\gamma$ and $\gamma\%$ with the two planned paired contrasts (Healthy $>$ SCZ; SCZ-speech $>$ SCZ-semantics).
Let the baseline external drive for utterance $u$ be $I^{(u)}_{\mathrm{ext}}(t)$. The nominal gains for the three conditions are
\begin{equation}
  g_H=1.00,\qquad g_S=0.75,\qquad g_{\mathrm{SEM}}=0.55.
  \label{eq:nominal-gains}
\end{equation}

\subsubsection{Sensitivity to multiplicative input-gain perturbations}

We examine multiplicative perturbations 
$j \in \{0.8, 0.9, 1.0, 1.1, 1.2\}$ under two schemes:
(1) \emph{Common-scale perturbation}, multiplying all three gains by $j$; 
and (2) \emph{Relative-scale perturbation}, keeping Healthy at $1.00$ and 
multiplying only the two SCZ gains by $j$.
Both can be written compactly as
\begin{equation}
  \tilde{I}^{(u)}_{\mathrm{ext},c}(t)=\tilde{g}_c(j)\,I^{(u)}_{\mathrm{ext}}(t),
  \label{eq:jitter}
\end{equation}
with $\tilde{g}_c(j)$ defined by the corresponding rule. Under identical network and noise settings, we integrate the model for each $j$ and utterance $u$, compute ERSP$_\gamma$ and $\gamma\%$, and perform the two planned paired $t$-tests.

\subsubsection{Equal-energy control (RMS pre-normalization)}

To remove confounds due to differences in total energy across utterances, we
first normalize each raw external drive $I_{\mathrm{ext}}^{(u)}(t)$ to a common
RMS and only then apply the condition gains. The procedure has two steps:

\begin{enumerate}
  \item \textit{Pre-normalize RMS.}  
  Define the dataset-level target RMS:
  \begin{equation}
    \mathrm{RMS}_{\text{target}}
    = \frac{1}{U}\sum_{u=1}^{U}\mathrm{RMS}\!\big(I_{\mathrm{ext}}^{(u)}\big).
    \label{eq:rmstarget}
  \end{equation}
  For each utterance $u$, set an amplitude factor
  \begin{equation}
    \alpha_u =
    \frac{\mathrm{RMS}_{\text{target}}}
         {\mathrm{RMS}\!\big(I_{\mathrm{ext}}^{(u)}\big)+\varepsilon},
    \qquad
    \tilde I_{\mathrm{ext}}^{(u)}(t)
      = \alpha_u\, I_{\mathrm{ext}}^{(u)}(t),
    \label{eq:alpha}
  \end{equation}
  where $\varepsilon>0$ is a small positive constant for numerical stability.

  \item \textit{Apply condition gains and re-estimate statistics.}  
  Apply the three baseline condition gains
  \[
    g_H = 1.00, \qquad
    g_S = 0.75, \qquad
    g_{\mathrm{SEM}} = 0.55,
  \]
\end{enumerate}

to the pre-normalized drive \(\tilde I_{\mathrm{ext}}^{(u)}(t)\), yielding the
condition-specific inputs
\begin{equation}
\begin{aligned}
  I^{\mathrm{H}}(t)   &= g_{\mathrm{H}}\,\tilde I_{\mathrm{ext}}^{(u)}(t),\\
  I^{\mathrm{S}}(t)   &= g_{\mathrm{S}}\,\tilde I_{\mathrm{ext}}^{(u)}(t),\\
  I^{\mathrm{SEM}}(t) &= g_{\mathrm{SEM}}\,\tilde I_{\mathrm{ext}}^{(u)}(t).
\end{aligned}
\label{eq:cond-inputs}
\end{equation}
Under identical network and noise settings, integrate the model for each
condition and utterance \(u\), compute the two primary metrics
\(\mathrm{ERSP}_\gamma\) and \(\gamma\%\), and run the two planned paired
\(t\)-tests (Healthy \(>\) SCZ; SCZ-speech \(>\) SCZ-semantics), reporting the
corresponding paired-samples effect size \(d_z\).

\section{Results}\label{sec:results}
\subsection{Verification of Network--Single-Node Equivalence}\label{sec:equiv}
\begin{figure*}[!htbp]
	\centering
	\begin{subfigure}[t]{0.48\textwidth}
		\centering
		\includegraphics[width=\linewidth]{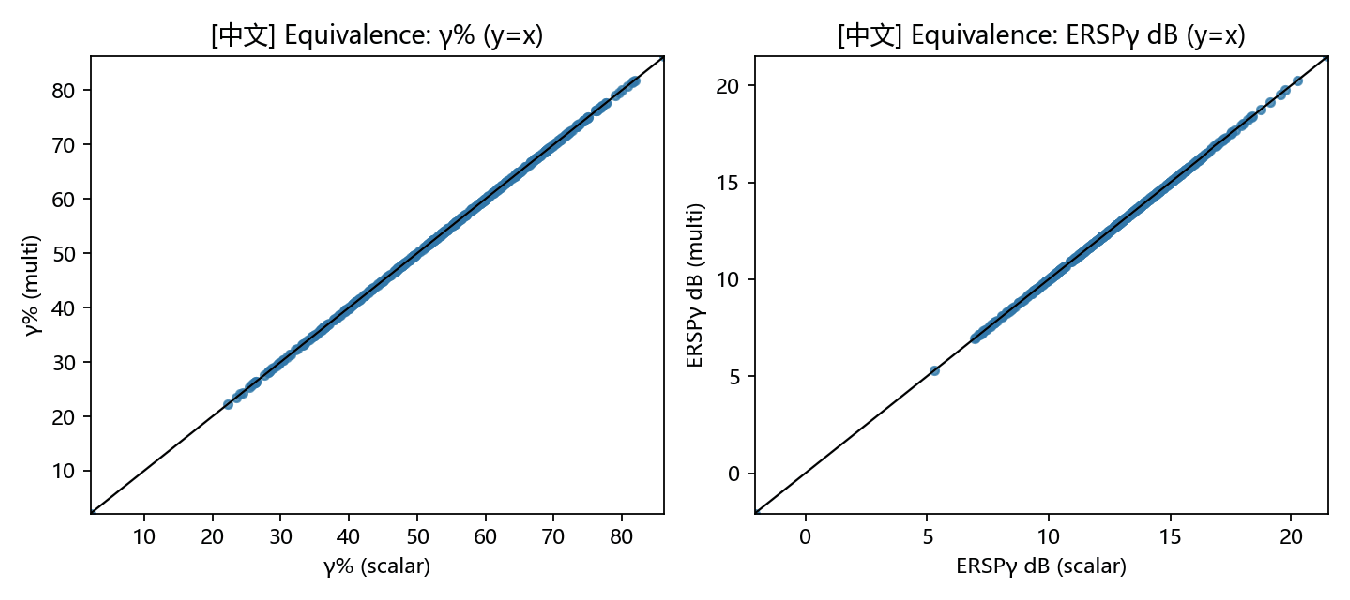}
		\caption{Mandarin TTS corpus}
		\label{fig:equiv-cn}
	\end{subfigure}\hfill
	\begin{subfigure}[t]{0.48\textwidth}
		\centering
		\includegraphics[width=\linewidth]{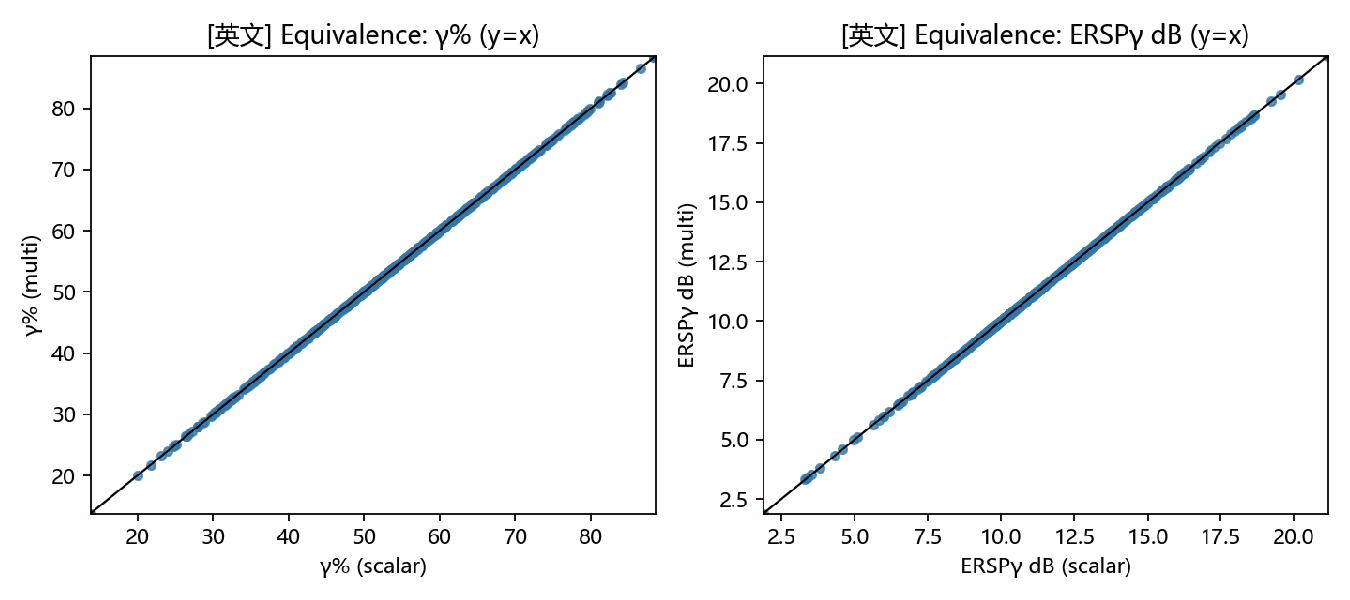}
		\caption{English TTS corpus}
		\label{fig:equiv-en}
	\end{subfigure}
	
	\vspace{0.3cm} 
	
	\begin{subfigure}[t]{0.48\textwidth}
		\centering
		\includegraphics[width=\linewidth]{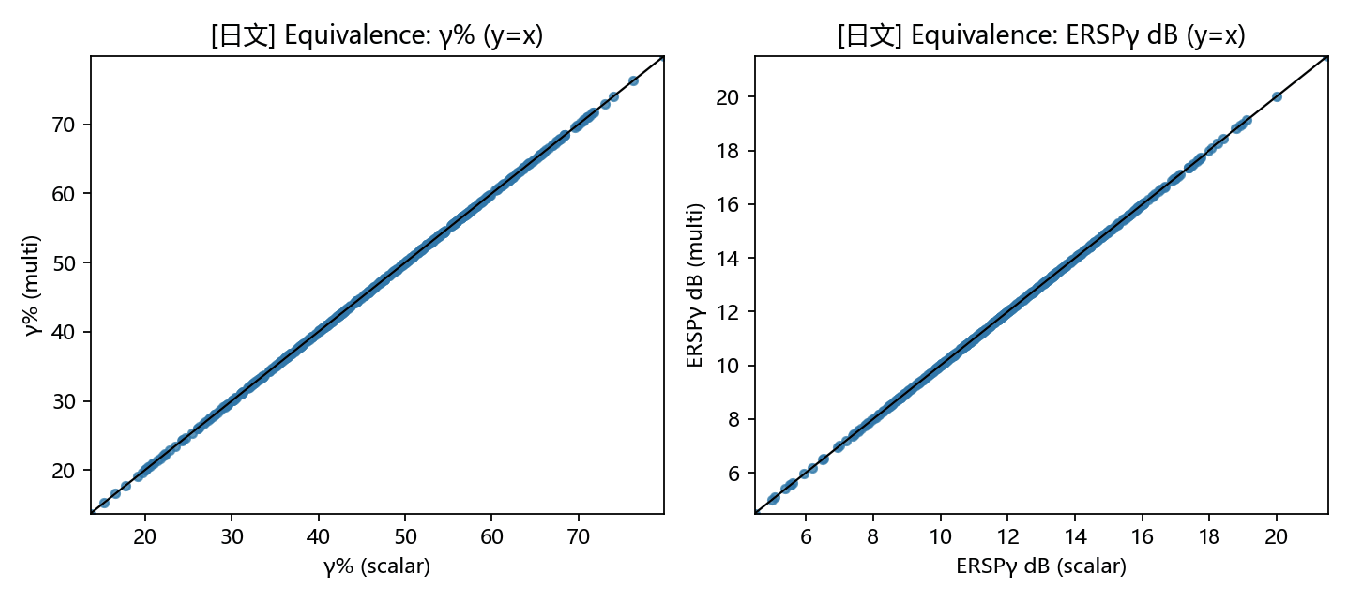}
		\caption{Japanese TTS corpus}
		\label{fig:equiv-ja}
	\end{subfigure}\hfill
	\begin{subfigure}[t]{0.48\textwidth}
		\centering
		\includegraphics[width=\linewidth]{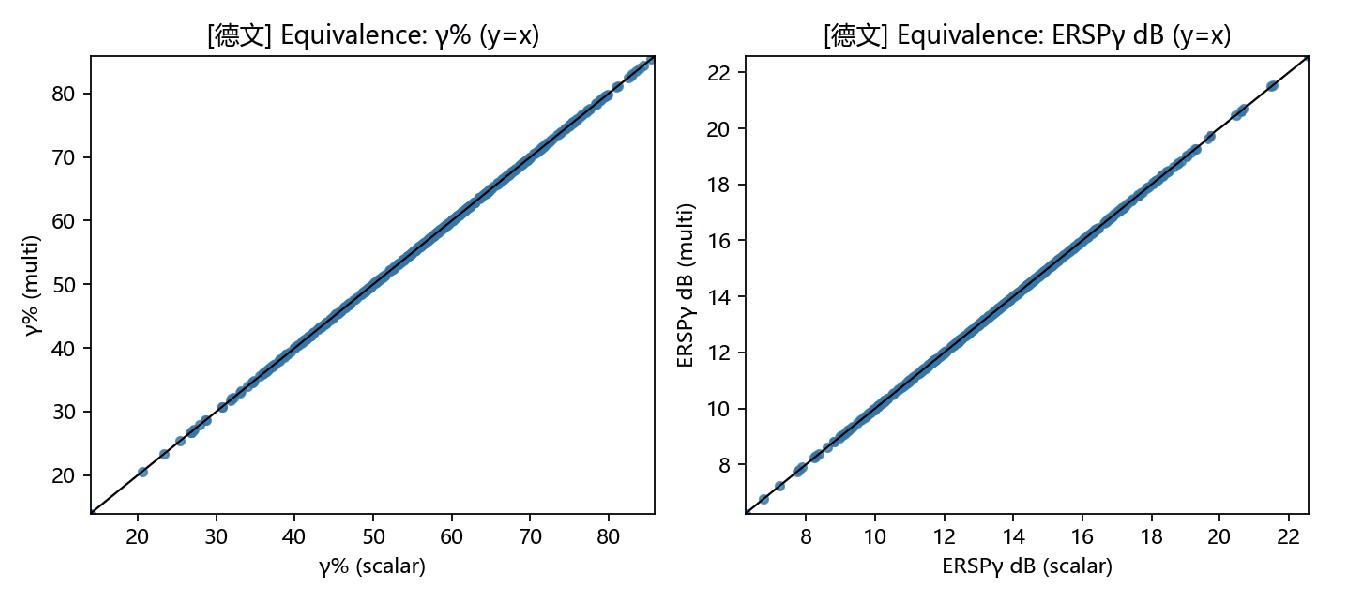}
		\caption{German TTS corpus}
		\label{fig:equiv-de}
	\end{subfigure}
	
	\vspace{0.3cm}
	
	\begin{subfigure}[t]{0.48\textwidth}
		\centering
		\includegraphics[width=\linewidth]{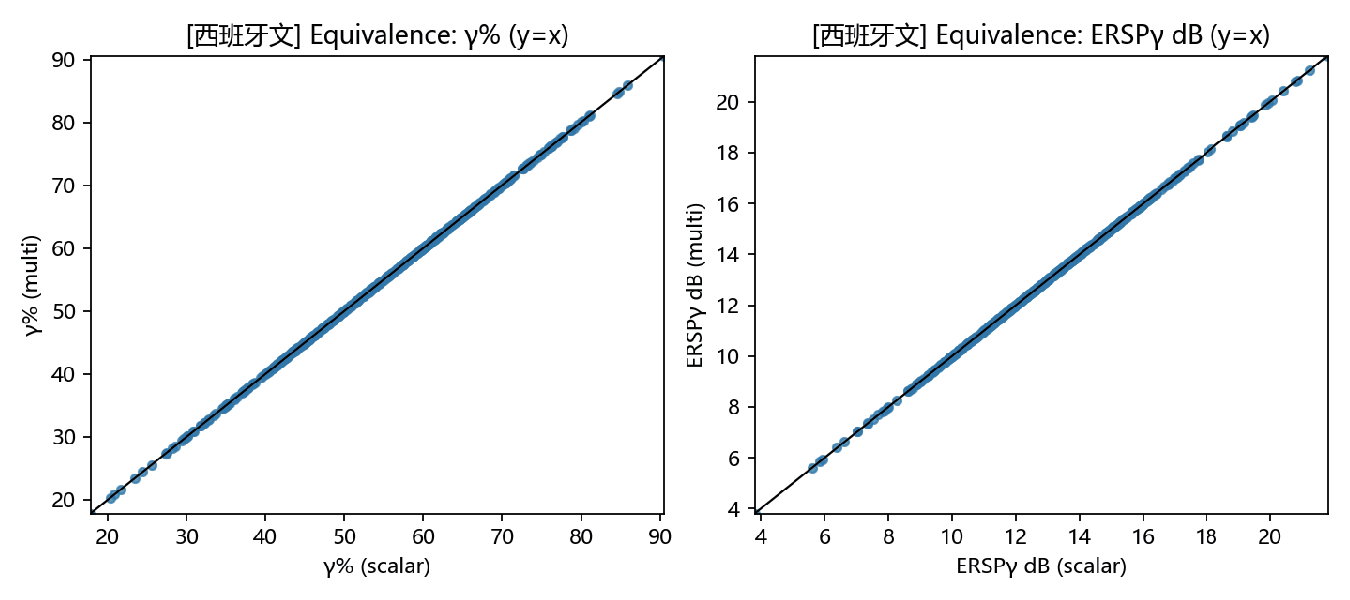}
		\caption{Spanish TTS corpus}
		\label{fig:equiv-es}
	\end{subfigure}\hfill
	\begin{subfigure}[t]{0.48\textwidth}
		\centering
		\includegraphics[width=\linewidth]{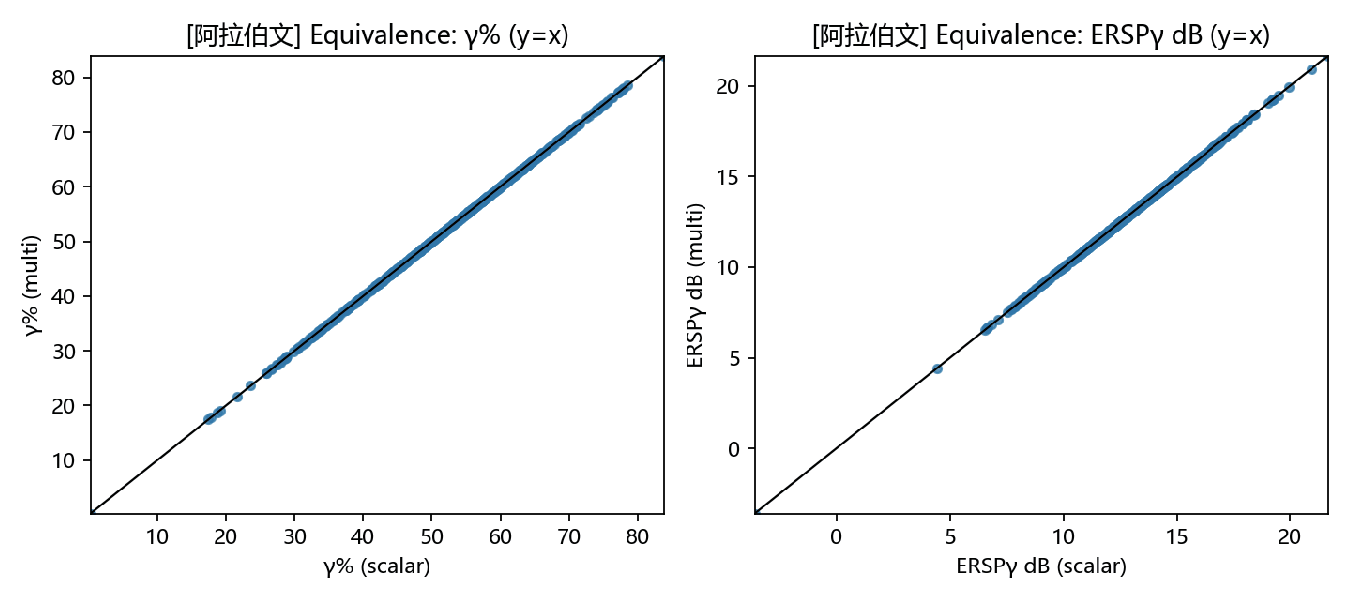}
		\caption{Arabic TTS corpus}
		\label{fig:equiv-ar}
	\end{subfigure}
	
	\caption{Metric equivalence between the multi-channel network and
		its single-node reduction across six languages.
		Each panel shows scatter plots of $\gamma\%$ and ERSP$_\gamma$ from
		the multi-channel model versus its single-node equivalent.
		Each dot corresponds to one utterance/segment; the solid line is
		$y=x$.
		Points lie essentially on the identity line in all languages,
		confirming numerical equivalence of the two formulations.}
	\label{fig:equiv}
\end{figure*}

Under fixed parameters and an identical external drive across channels,
the multi-channel coupled network we constructed is strictly
synchronized.
If each coupling matrix has rows that sum to the prescribed target
weight $\omega_{XY}$ and every channel receives the same input
$I_{\mathrm{ext}}(t)$, then each channel’s states evolve according to
equations that are exactly equivalent to those of a single-node
Wilson--Cowan system with the same effective row-sum weights.
Consequently, the spatial mean of the excitatory population, i.e.,
the local field potential,
\begin{equation}
	\mathrm{LFP}(t)=\frac{1}{C}\sum_{i=1}^{C} \tau^{(i)}_{E}(t),
\end{equation}
is theoretically identical to the single-node trajectory
$\tau_{E}(t)$.
We verified this claim numerically across all six languages.
For every stimulus and condition in the Mandarin, English, Japanese, 
German, Spanish, and Arabic corpora, we simulated a 64-channel 
network model and its single-node equivalent under identical 
initial states and matched noise sequences.
For each realization we computed the norms
$\|\mathrm{LFP}(t)-\tau_{E}(t)\|_{\infty}$ and
$\|\mathrm{LFP}(t)-\tau_{E}(t)\|_{2}$.
In all cases the time series from the two models were indistinguishable
within numerical precision, with maximal absolute errors below
$10^{-12}$ (Fig.~\ref{fig:equiv}).
The corresponding ERSP$_{\gamma}$ and $\gamma\%$ metrics coincide at
both per-utterance and group levels; any residual differences arise
only from floating-point round-off.

This validation justifies implementing subsequent stability analyses,
parameter-sensitivity tests, and pharmacological perturbations using
the single-node reduction in place of the full network.
The reduction substantially reduces computational cost while preserving
the dynamical behavior of interest across diverse linguistic prosodies 
and providing additional numerical support for the theoretical equivalence.

\begin{figure*}[!htbp]
	\centering
	\begin{subfigure}[t]{0.48\textwidth}
		\centering
		\includegraphics[width=\linewidth]{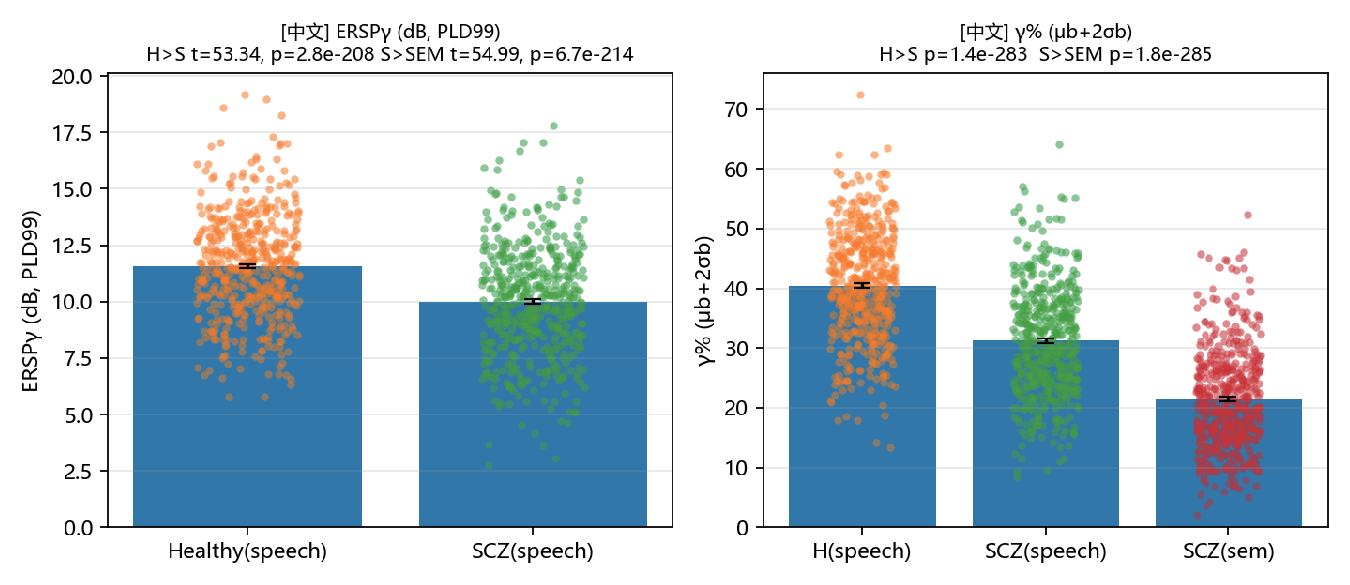}
		\caption{Mandarin TTS corpus (500 segments)}
		\label{fig:comparison-cn}
	\end{subfigure}\hfill
	\begin{subfigure}[t]{0.48\textwidth}
		\centering
		\includegraphics[width=\linewidth]{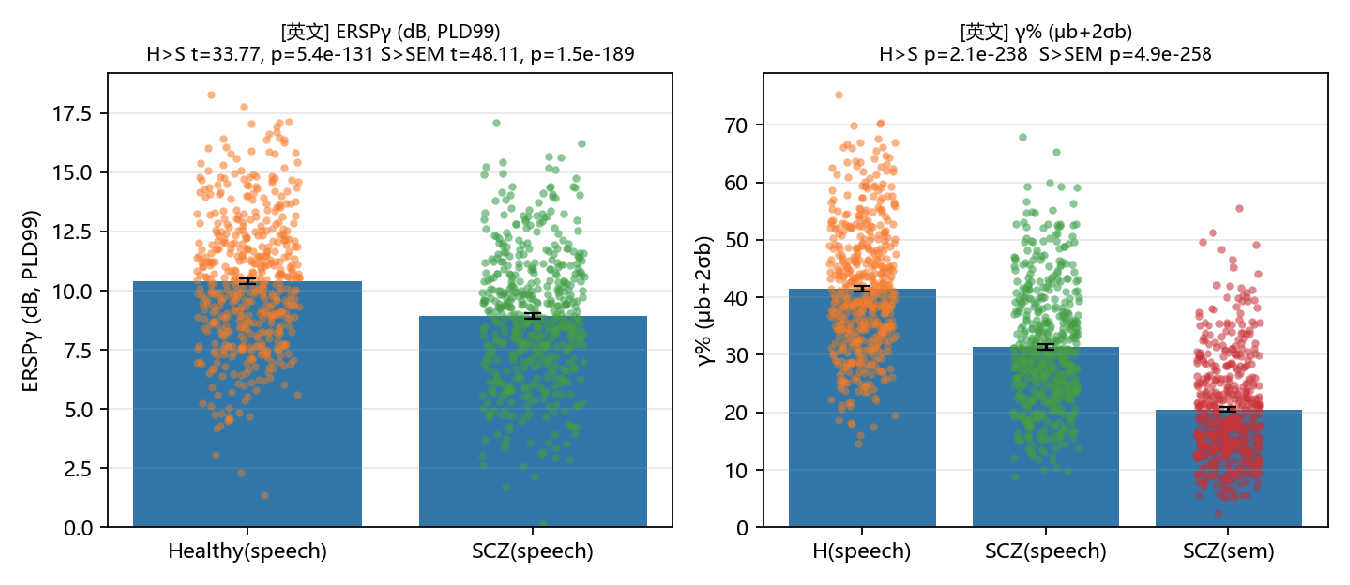}
		\caption{English TTS corpus (500 segments)}
		\label{fig:comparison-en}
	\end{subfigure}
	
	\vspace{0.3cm}
	
	\begin{subfigure}[t]{0.48\textwidth}
		\centering
		\includegraphics[width=\linewidth]{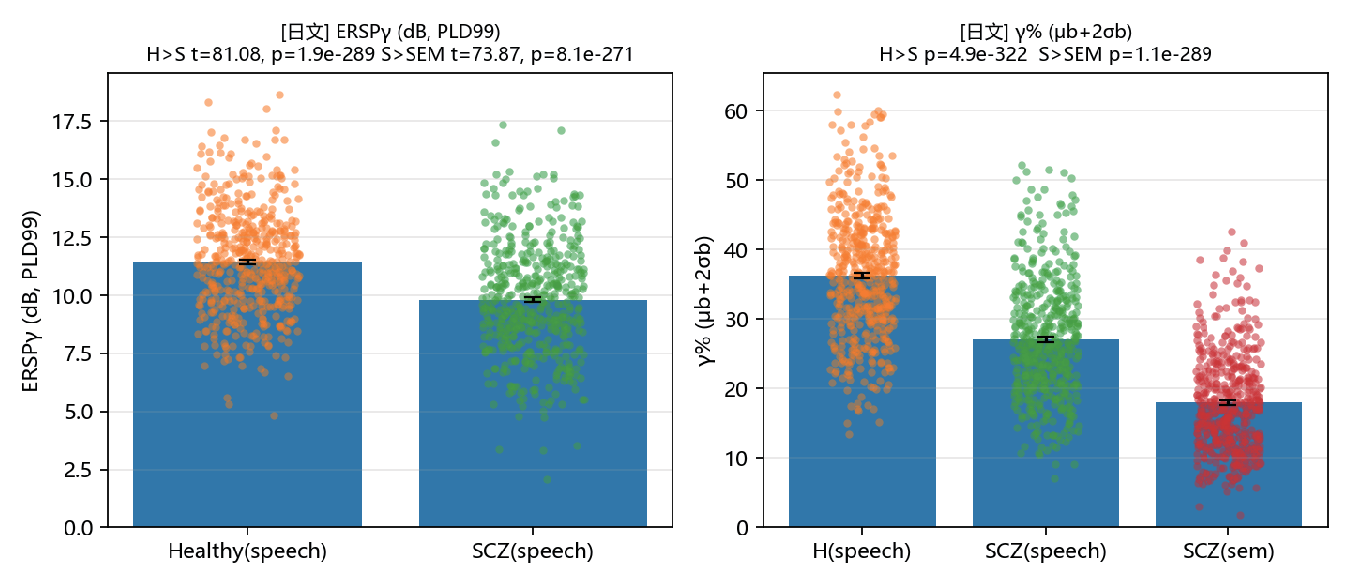}
		\caption{Japanese TTS corpus (500 segments)}
		\label{fig:comparison-jp}
	\end{subfigure}\hfill
	\begin{subfigure}[t]{0.48\textwidth}
		\centering
		\includegraphics[width=\linewidth]{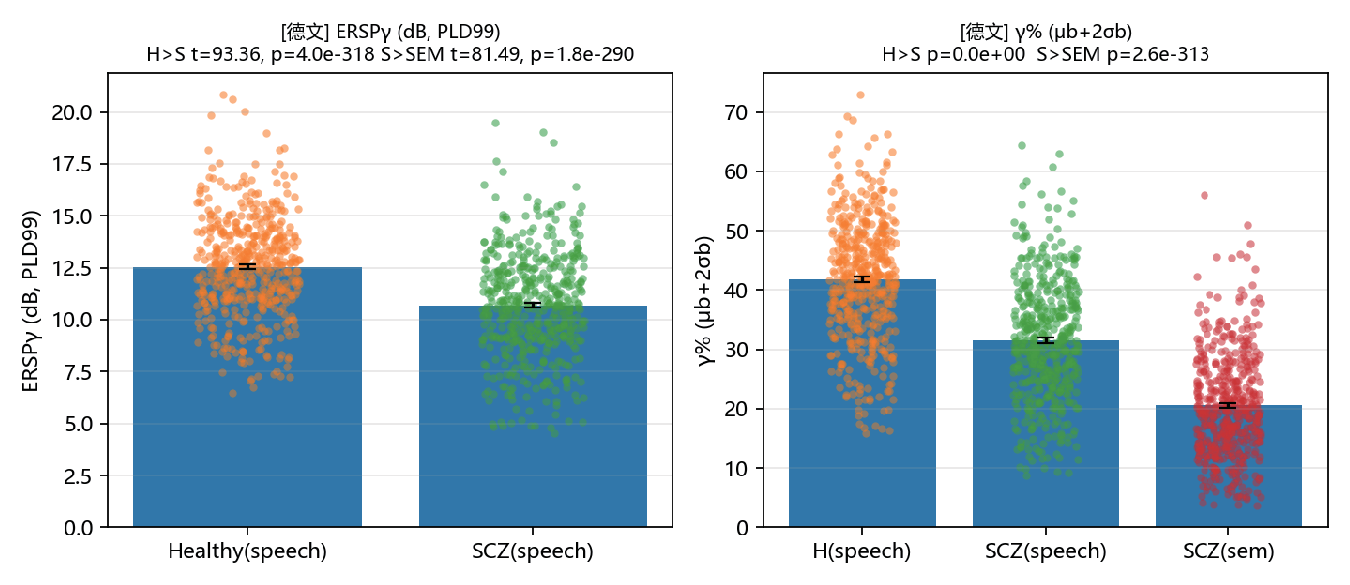}
		\caption{German TTS corpus (500 segments)}
		\label{fig:comparison-de}
	\end{subfigure}
	
	\vspace{0.3cm}
	
	\begin{subfigure}[t]{0.48\textwidth}
		\centering
		\includegraphics[width=\linewidth]{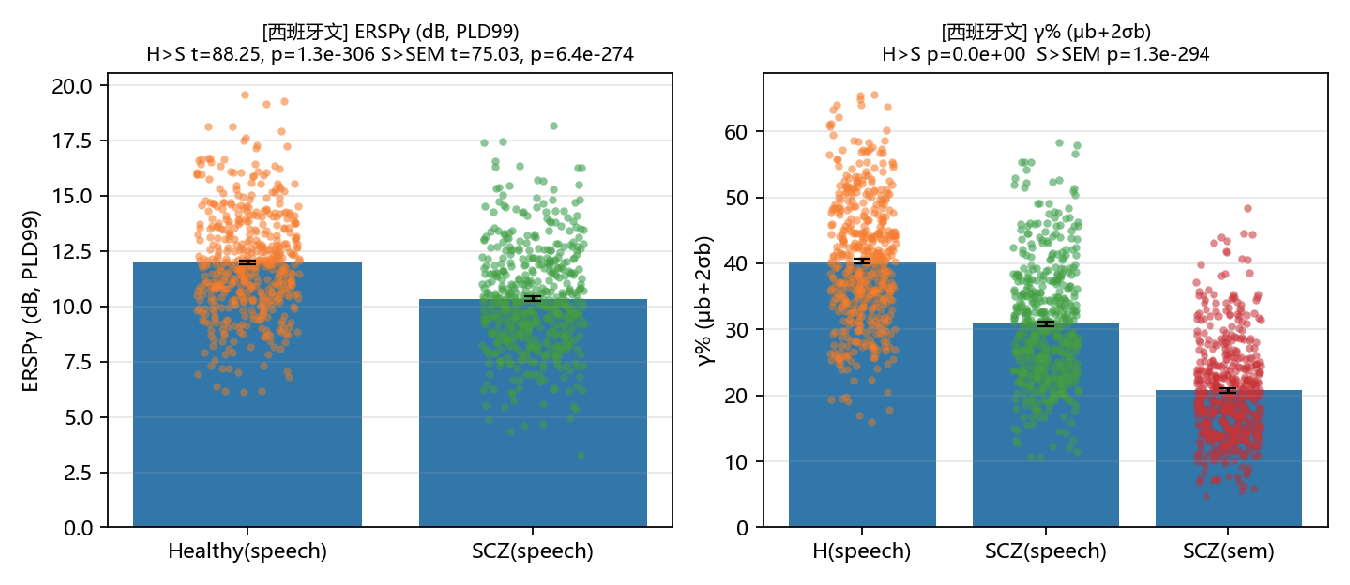}
		\caption{Spanish TTS corpus (500 segments)}
		\label{fig:comparison-es}
	\end{subfigure}\hfill
	\begin{subfigure}[t]{0.48\textwidth}
		\centering
		\includegraphics[width=\linewidth]{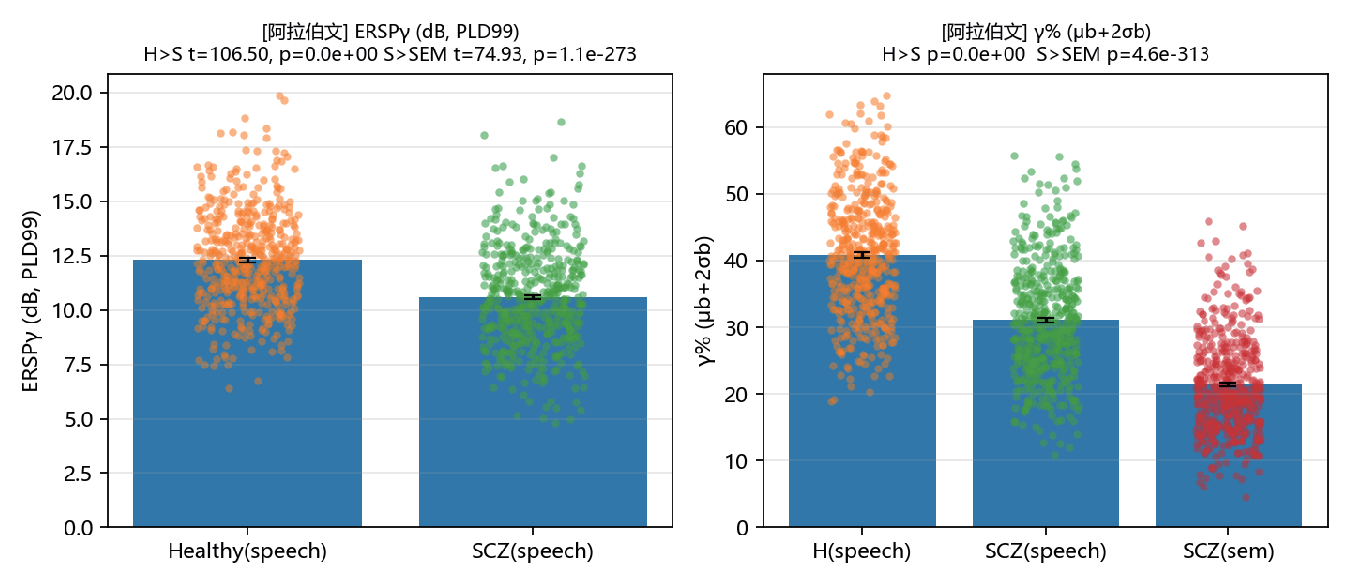}
		\caption{Arabic TTS corpus (500 segments)}
		\label{fig:comparison-ar}
	\end{subfigure}
	
	\caption{Comparison of primary metrics across experimental conditions
		for six languages.
		Each panel shows ERSP$_\gamma$ in dB (left subplot) and
		$\gamma\%$ (right subplot) for Healthy (speech/semantics), SCZ (speech) and
		SCZ (semantics).
		Bars indicate means with s.e.m.\ and dots show individual
		utterances/segments.
		In all tested languages (Mandarin, English, Japanese, German, Spanish, 
		and Arabic), the ordering Healthy(speech/semantics) $>$ SCZ(speech) $>$ 
		SCZ(semantics) is strictly preserved.}
	\label{fig:comparison}
\end{figure*}

\subsection{Comparison of primary metrics}\label{subsec:primary-metrics}

We first evaluated how well the model reproduces the group ordering of
$\gamma$-band activity across experimental conditions using two
complementary outcome measures: ERSP$_\gamma$ in dB and the modulation
depth measure $\gamma\%$.
Here we used six matched text-to-speech (TTS) corpora, each containing 
500 segments: Mandarin, English, Japanese, German, Spanish, and Arabic. 
For each language, we considered three conditions:
Healthy (speech/semantics), SCZ (speech), and SCZ (semantics)
(abbrev., H, S, SEM).
Across all six languages, the two metrics yield the same graded ordering
(Fig.~\ref{fig:comparison}):
\emph{Healthy(speech/semantics) $>$ SCZ(speech) $>$ SCZ(semantics)}.
Within each language, subject-wise paired comparisons show that most
utterances/segments increase jointly on both metrics when moving from
SEM to S and from S to H.
The planned contrasts H $>$ S and S $>$ SEM are highly significant for
both ERSP$_\gamma$ and $\gamma\%$ across all tested languages
(paired $t$-tests with extremely small $p$-values; see
Table~\ref{tab:primary-metrics} for six languages examples).
These main effects are consistent with mechanistic interpretations in
which speech-driven input places the Healthy E/I network in a regime
more prone to $\gamma$ oscillations than in the SCZ conditions.

\begin{table*}[!htbp]
	\centering
	\scriptsize
	\caption{Comparison of primary $\gamma$-band metrics across six TTS corpora ($n = 500$ segments per language; paired $t$-tests). Data representing mean $\pm$ SD, $t$-statistics, and $p$-values.}
	\label{tab:primary-metrics}
	\resizebox{\textwidth}{!}{%
		\begin{tabular}{lllcccc}
			\toprule
			Language & Metric & Comparison & Cond. A (mean$\pm$SD) & Cond. B (mean$\pm$SD) & $t$ & $p$ \\
			\midrule
			\multirow{4}{*}{Mandarin} & ERSP$_\gamma$ (dB) & H $>$ S & $11.57 \pm 2.27$ & $10.00 \pm 2.39$ & 53.34 & $2.76 \times 10^{-208}$ \\
			& ERSP$_\gamma$ (dB) & S $>$ SEM & $10.00 \pm 2.39$ & $7.95 \pm 2.69$ & 54.99 & $6.68 \times 10^{-214}$ \\
			& $\gamma\%$ (thr) & H $>$ S & $40.42 \pm 9.11$ & $31.24 \pm 9.15$ & 78.75 & $1.39 \times 10^{-283}$ \\
			& $\gamma\%$ (thr) & S $>$ SEM & $31.24 \pm 9.15$ & $21.51 \pm 8.25$ & 79.49 & $1.84 \times 10^{-285}$ \\
			\midrule
			\multirow{4}{*}{English} & ERSP$_\gamma$ (dB) & H $>$ S & $10.13 \pm 2.81$ & $8.71 \pm 2.84$ & 30.79 & $1.98 \times 10^{-117}$ \\
			& ERSP$_\gamma$ (dB) & S $>$ SEM & $8.71 \pm 2.84$ & $6.67 \pm 2.84$ & 42.55 & $3.75 \times 10^{-168}$ \\
			& $\gamma\%$ (thr) & H $>$ S & $41.57 \pm 11.01$ & $31.36 \pm 10.63$ & 62.58 & $2.11 \times 10^{-238}$ \\
			& $\gamma\%$ (thr) & S $>$ SEM & $31.36 \pm 10.63$ & $20.46 \pm 9.01$ & 69.24 & $4.89 \times 10^{-258}$ \\
			\midrule
			\multirow{4}{*}{Japanese} & ERSP$_\gamma$ (dB) & H $>$ S & $11.45 \pm 2.27$ & $9.81 \pm 2.40$ & 81.08 & $1.91 \times 10^{-289}$ \\
			& ERSP$_\gamma$ (dB) & S $>$ SEM & $9.81 \pm 2.40$ & $7.73 \pm 2.59$ & 73.87 & $8.13 \times 10^{-271}$ \\
			& $\gamma\%$ (thr) & H $>$ S & $36.29 \pm 9.00$ & $27.11 \pm 8.48$ & 95.16 & $4.94 \times 10^{-322}$ \\
			& $\gamma\%$ (thr) & S $>$ SEM & $27.11 \pm 8.48$ & $17.99 \pm 7.05$ & 81.17 & $1.10 \times 10^{-289}$ \\
			\midrule
			\multirow{4}{*}{German} & ERSP$_\gamma$ (dB) & H $>$ S & $12.55 \pm 2.37$ & $10.69 \pm 2.51$ & 93.36 & $3.97 \times 10^{-318}$ \\
			& ERSP$_\gamma$ (dB) & S $>$ SEM & $10.69 \pm 2.51$ & $8.29 \pm 2.77$ & 81.49 & $1.77 \times 10^{-290}$ \\
			& $\gamma\%$ (thr) & H $>$ S & $41.79 \pm 9.97$ & $31.50 \pm 9.96$ & 104.16 & $< 1 \times 10^{-323}$ \\
			& $\gamma\%$ (thr) & S $>$ SEM & $31.50 \pm 9.96$ & $20.58 \pm 8.66$ & 91.19 & $2.63 \times 10^{-313}$ \\
			\midrule
			\multirow{4}{*}{Spanish} & ERSP$_\gamma$ (dB) & H $>$ S & $11.98 \pm 2.31$ & $10.33 \pm 2.35$ & 88.25 & $1.27 \times 10^{-306}$ \\
			& ERSP$_\gamma$ (dB) & S $>$ SEM & $10.33 \pm 2.35$ & $8.22 \pm 2.49$ & 75.03 & $6.36 \times 10^{-274}$ \\
			& $\gamma\%$ (thr) & H $>$ S & $40.34 \pm 9.37$ & $30.84 \pm 8.98$ & 97.97 & $< 1 \times 10^{-323}$ \\
			& $\gamma\%$ (thr) & S $>$ SEM & $30.84 \pm 8.98$ & $20.78 \pm 7.51$ & 83.18 & $1.28 \times 10^{-294}$ \\
			\midrule
			\multirow{4}{*}{Arabic} & ERSP$_\gamma$ (dB) & H $>$ S & $12.30 \pm 2.23$ & $10.59 \pm 2.27$ & 106.50 & $< 1 \times 10^{-323}$ \\
			& ERSP$_\gamma$ (dB) & S $>$ SEM & $10.59 \pm 2.27$ & $8.53 \pm 2.45$ & 74.93 & $1.14 \times 10^{-273}$ \\
			& $\gamma\%$ (thr) & H $>$ S & $40.84 \pm 8.89$ & $31.09 \pm 8.42$ & 108.72 & $< 1 \times 10^{-323}$ \\
			& $\gamma\%$ (thr) & S $>$ SEM & $31.09 \pm 8.42$ & $21.41 \pm 7.14$ & 91.08 & $4.63 \times 10^{-313}$ \\
			\bottomrule
		\end{tabular}%
	}
\end{table*}

\subsection{Equal-energy control}\label{subsec:equal-energy}

\begin{figure*}[!htbp]
	\centering
	\begin{subfigure}[t]{0.48\textwidth}
		\centering
		\includegraphics[width=\linewidth]{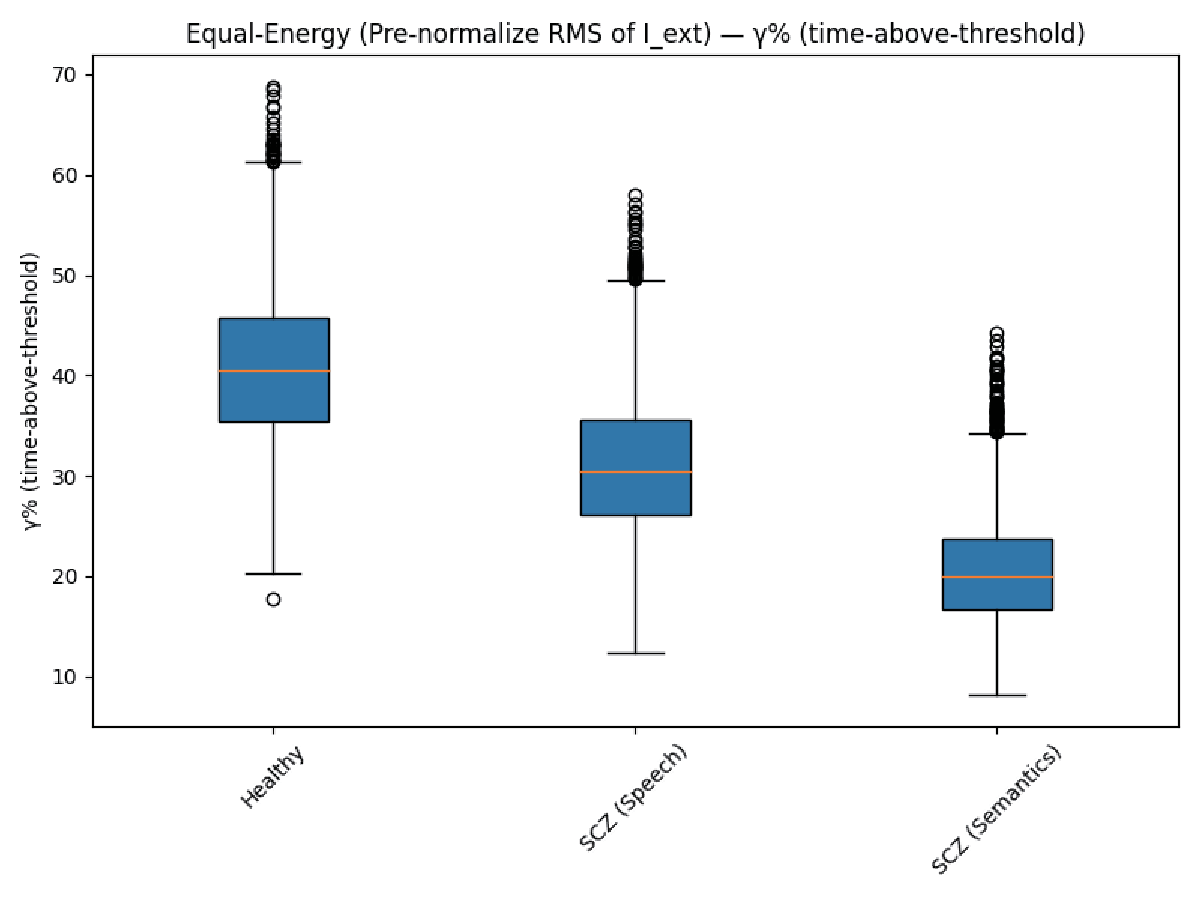}
		\caption{$\gamma\%$ (time-above-threshold) across all languages}
		\label{fig:equal-energy-gamma}
	\end{subfigure}\hfill
	\begin{subfigure}[t]{0.48\textwidth}
		\centering
		\includegraphics[width=\linewidth]{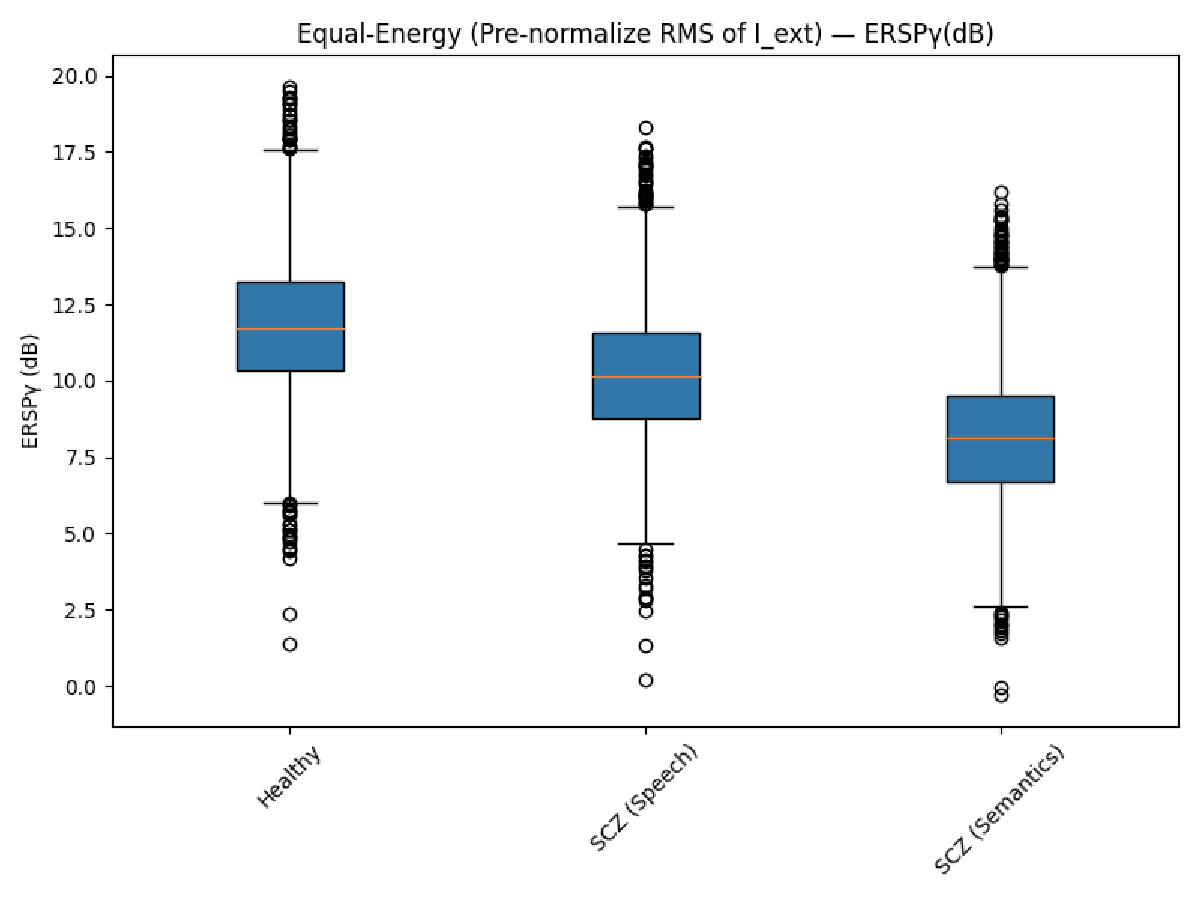}
		\caption{ERSP$_\gamma$ (dB) across all languages}
		\label{fig:equal-energy-ersp}
	\end{subfigure}
	
	\caption{Equal-energy control across conditions for six languages.
		For each stimulus, the external drive was RMS-normalized before
		simulation.
		Left subpanel: $\gamma\%$ (time-above-threshold) for each condition. 
		Right subpanel: ERSP$_\gamma$ in dB for each condition. 
		Planned contrasts Healthy $>$ SCZ and SCZ(speech) $>$ SCZ(semantics) remain highly significant across 
		all tested languages, indicating that the ordering is independent 
		of raw input energy.}
	\label{fig:equal-energy-combined}
\end{figure*}

Differences in total input energy could in principle confound the
comparison across conditions.
To rule out this possibility, we repeated the simulations after
normalizing each stimulus to a common root-mean-square (RMS) amplitude
before constructing the external drive $I_{\mathrm{ext}}(t)$.
For all six languages (Mandarin, English, Japanese, German, Spanish, 
and Arabic), the equal-energy analysis preserves the main
ordering in both outcome measures
(Fig.~\ref{fig:equal-energy-combined} and Table~\ref{tab:equal-energy}):
Healthy(speech/semantics) remains above SCZ(speech), and SCZ(speech) remains
above SCZ(semantics) for both $\gamma\%$ and ERSP$_\gamma$.
The planned paired contrasts H $>$ S and S $>$ SEM remain highly
significant in each language, confirming the robustness of the effect 
across diverse phonological structures.
Thus, the main effect does not reflect trivial consequences of input
energy but rather systematic differences in how network gain control
engages E/I circuitry under the three conditions.

\begin{table*}[!htbp]
	\centering
	\scriptsize
	\caption{Paired contrasts after equal-energy normalization across six TTS corpora ($n = 500$ segments per language). Each stimulus was RMS-normalized before simulation to ensure input energy consistency.}
	\label{tab:equal-energy}
	\resizebox{\textwidth}{!}{%
		\begin{tabular}{lllcccc}
			\toprule
			Language & Metric & Contrast & Cond. A (mean$\pm$SD) & Cond. B (mean$\pm$SD) & $t$ & $p$ \\
			\midrule
			\multirow{4}{*}{Mandarin} & $\gamma\%$ (equal-energy) & H $>$ S & $41.14 \pm 7.21$ & $31.95 \pm 6.85$ & 85.62 & $1.80 \times 10^{-300}$ \\
			& $\gamma\%$ (equal-energy) & S $>$ SEM & $31.95 \pm 6.85$ & $21.66 \pm 5.58$ & 79.35 & $4.01 \times 10^{-285}$ \\
			& ERSP$_\gamma$ (dB, equal-energy) & H $>$ S & $11.70 \pm 2.08$ & $10.18 \pm 2.07$ & 58.22 & $1.25 \times 10^{-224}$ \\
			& ERSP$_\gamma$ (dB, equal-energy) & S $>$ SEM & $10.18 \pm 2.07$ & $8.23 \pm 2.14$ & 61.90 & $2.69 \times 10^{-236}$ \\
			\midrule
			\multirow{4}{*}{English} & $\gamma\%$ (equal-energy) & H $>$ S & $42.30 \pm 8.61$ & $32.20 \pm 7.65$ & 66.35 & $1.06 \times 10^{-249}$ \\
			& $\gamma\%$ (equal-energy) & S $>$ SEM & $32.20 \pm 7.65$ & $20.66 \pm 5.79$ & 75.61 & $1.89 \times 10^{-275}$ \\
			& ERSP$_\gamma$ (dB, equal-energy) & H $>$ S & $10.48 \pm 2.64$ & $9.16 \pm 2.47$ & 30.82 & $1.37 \times 10^{-117}$ \\
			& ERSP$_\gamma$ (dB, equal-energy) & S $>$ SEM & $9.16 \pm 2.47$ & $7.17 \pm 2.37$ & 50.81 & $2.18 \times 10^{-199}$ \\
			\midrule
			\multirow{4}{*}{Japanese} & $\gamma\%$ (equal-energy) & H $>$ S & $36.74 \pm 7.04$ & $27.36 \pm 6.40$ & 96.99 & $< 1 \times 10^{-323}$ \\
			& $\gamma\%$ (equal-energy) & S $>$ SEM & $27.36 \pm 6.40$ & $17.94 \pm 5.02$ & 85.74 & $9.13 \times 10^{-301}$ \\
			& ERSP$_\gamma$ (dB, equal-energy) & H $>$ S & $11.55 \pm 2.05$ & $9.95 \pm 2.09$ & 84.31 & $2.44 \times 10^{-297}$ \\
			& ERSP$_\gamma$ (dB, equal-energy) & S $>$ SEM & $9.95 \pm 2.09$ & $7.91 \pm 2.16$ & 87.97 & $5.51 \times 10^{-306}$ \\
			\midrule
			\multirow{4}{*}{German} & $\gamma\%$ (equal-energy) & H $>$ S & $42.64 \pm 7.19$ & $32.12 \pm 6.72$ & 111.61 & $< 1 \times 10^{-323}$ \\
			& $\gamma\%$ (equal-energy) & S $>$ SEM & $32.12 \pm 6.72$ & $20.64 \pm 5.44$ & 102.35 & $< 1 \times 10^{-323}$ \\
			& ERSP$_\gamma$ (dB, equal-energy) & H $>$ S & $12.72 \pm 2.07$ & $10.92 \pm 2.08$ & 117.64 & $< 1 \times 10^{-323}$ \\
			& ERSP$_\gamma$ (dB, equal-energy) & S $>$ SEM & $10.92 \pm 2.08$ & $8.58 \pm 2.08$ & 100.50 & $< 1 \times 10^{-323}$ \\
			\midrule
			\multirow{4}{*}{Spanish} & $\gamma\%$ (equal-energy) & H $>$ S & $40.85 \pm 7.81$ & $31.17 \pm 7.15$ & 99.77 & $< 1 \times 10^{-323}$ \\
			& $\gamma\%$ (equal-energy) & S $>$ SEM & $31.17 \pm 7.15$ & $20.97 \pm 5.50$ & 87.96 & $5.94 \times 10^{-306}$ \\
			& ERSP$_\gamma$ (dB, equal-energy) & H $>$ S & $12.07 \pm 2.14$ & $10.44 \pm 2.11$ & 98.71 & $< 1 \times 10^{-323}$ \\
			& ERSP$_\gamma$ (dB, equal-energy) & S $>$ SEM & $10.44 \pm 2.11$ & $8.42 \pm 2.09$ & 82.39 & $1.08 \times 10^{-292}$ \\
			\midrule
			\multirow{4}{*}{Arabic} & $\gamma\%$ (equal-energy) & H $>$ S & $41.17 \pm 7.10$ & $31.36 \pm 6.57$ & 106.32 & $< 1 \times 10^{-323}$ \\
			& $\gamma\%$ (equal-energy) & S $>$ SEM & $31.36 \pm 6.57$ & $21.54 \pm 5.37$ & 96.38 & $< 1 \times 10^{-323}$ \\
			& ERSP$_\gamma$ (dB, equal-energy) & H $>$ S & $12.37 \pm 2.07$ & $10.71 \pm 2.06$ & 109.34 & $< 1 \times 10^{-323}$ \\
			& ERSP$_\gamma$ (dB, equal-energy) & S $>$ SEM & $10.71 \pm 2.06$ & $8.72 \pm 2.06$ & 95.86 & $1.98 \times 10^{-323}$ \\
			\bottomrule
		\end{tabular}%
	}
\end{table*}

\subsection{Sensitivity to perturbations of input gain}\label{subsec:gain-sensitivity}

\begin{figure*}[!htbp]
	\centering
	\begin{subfigure}[t]{0.48\textwidth}
		\centering
		\includegraphics[width=\linewidth]{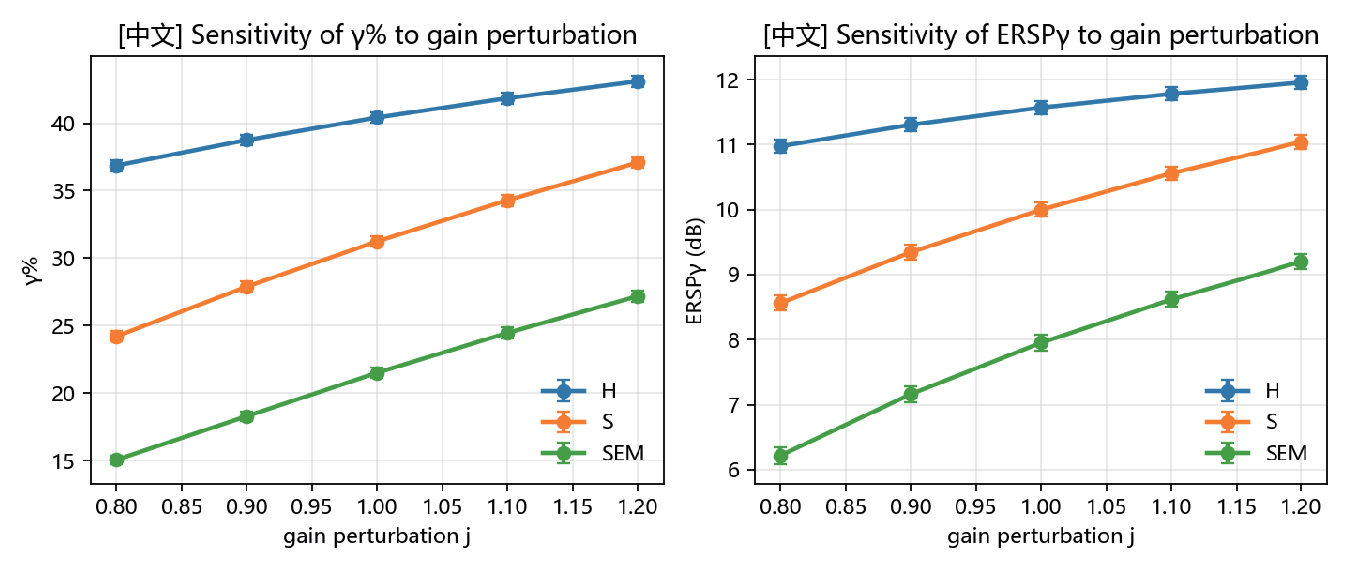}
		\caption{Mandarin TTS corpus}
		\label{fig:sens-cn}
	\end{subfigure}\hfill
	\begin{subfigure}[t]{0.48\textwidth}
		\centering
		\includegraphics[width=\linewidth]{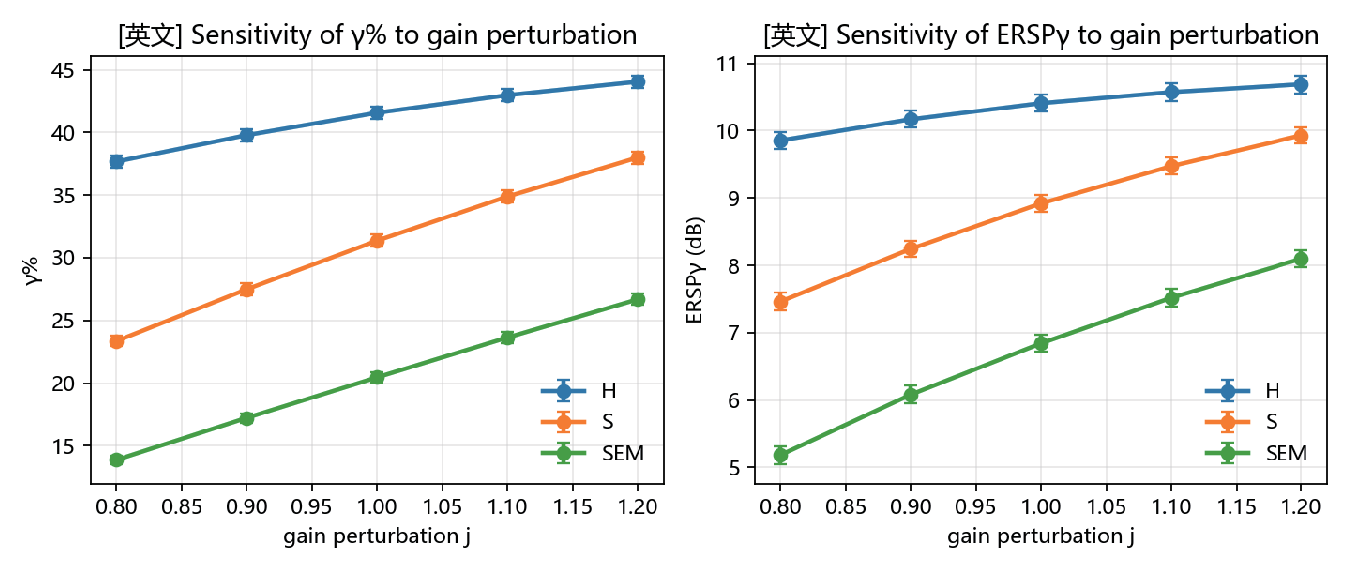}
		\caption{English TTS corpus}
		\label{fig:sens-en}
	\end{subfigure}
	
	\vspace{0.3cm}
	
	\begin{subfigure}[t]{0.48\textwidth}
		\centering
		\includegraphics[width=\linewidth]{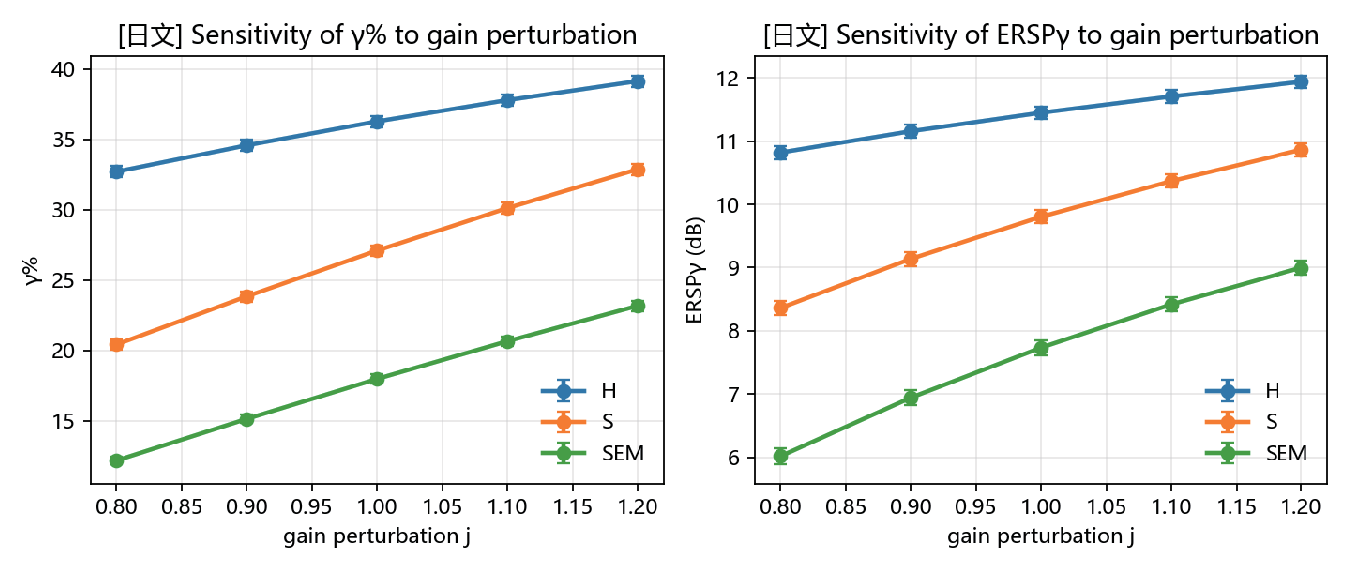}
		\caption{Japanese TTS corpus}
		\label{fig:sens-jp}
	\end{subfigure}\hfill
	\begin{subfigure}[t]{0.48\textwidth}
		\centering
		\includegraphics[width=\linewidth]{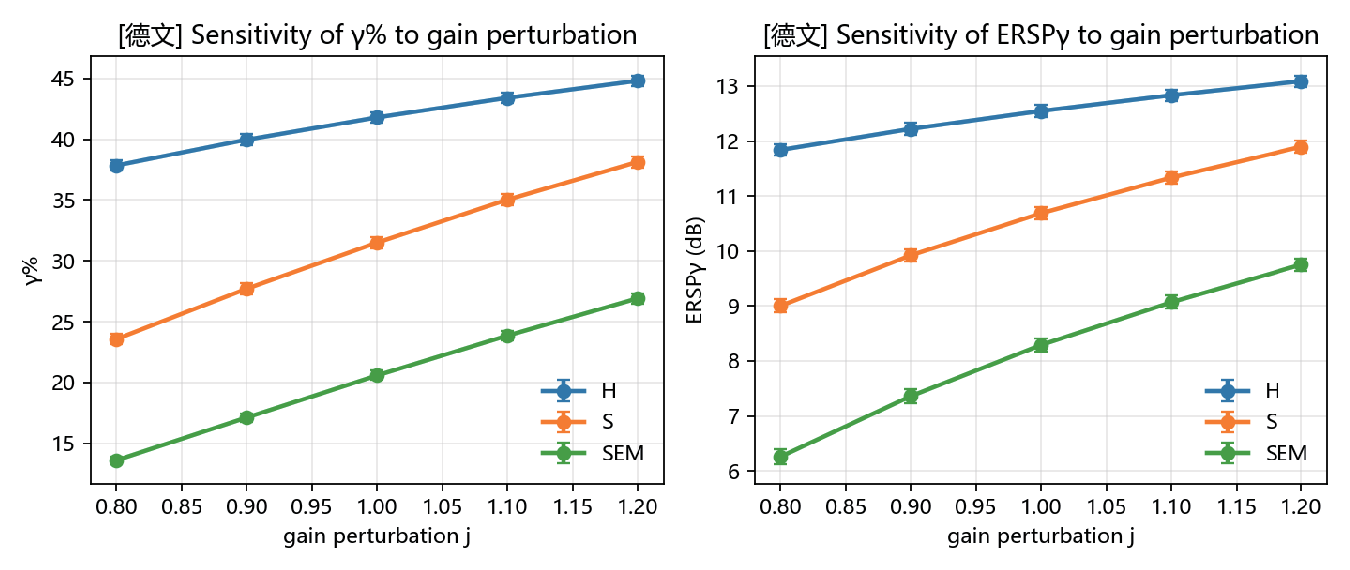}
		\caption{German TTS corpus}
		\label{fig:sens-de}
	\end{subfigure}
	
	\vspace{0.3cm}
	
	\begin{subfigure}[t]{0.48\textwidth}
		\centering
		\includegraphics[width=\linewidth]{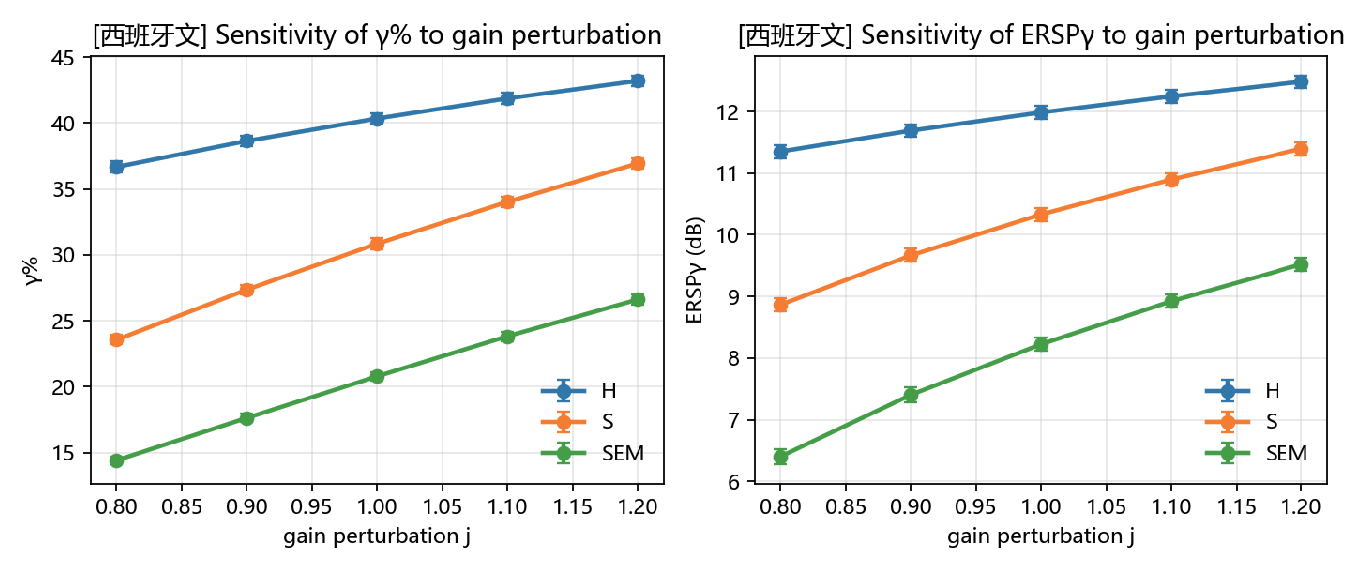}
		\caption{Spanish TTS corpus}
		\label{fig:sens-es}
	\end{subfigure}\hfill
	\begin{subfigure}[t]{0.48\textwidth}
		\centering
		\includegraphics[width=\linewidth]{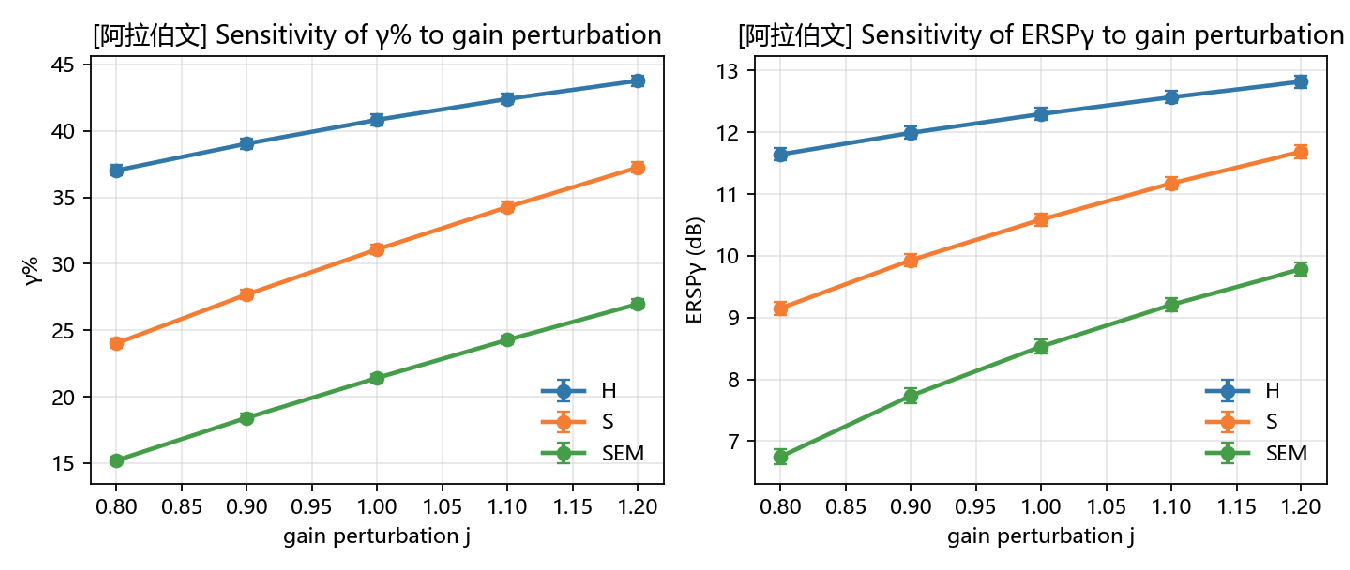}
		\caption{Arabic TTS corpus}
		\label{fig:sens-ar}
	\end{subfigure}
	
	\caption{Sensitivity of $\gamma\%$ and ERSP$_\gamma$ to
		multiplicative perturbations of input gain for six languages.
		Each panel shows mean\,$\pm$\,s.e.m.\ across utterances/segments as
		a function of the gain factor $j$ under both the common-scale and
		relative-scale schemes.
		Across all $j$ and in all languages, the ordering
		Healthy(speech/semantics) $>$ SCZ(speech) $>$ SCZ(semantics) is preserved.}
	\label{fig:sensitivities}
\end{figure*}

We next assessed the robustness of these conclusions to uncertainty in
input amplitude.
For each language, we implemented the same two perturbation schemes as
in the preregistered analysis: a global
(common-scale) perturbation in which all conditions share a common
multiplicative factor $j$, and a relative-scale perturbation in which
Healthy is fixed at $j=1$ while the SCZ conditions are scaled.
In both schemes we scanned gain factors
$j\in\{0.8,0.9,1.0,1.1,1.2\}$.
For all six languages (Mandarin, English, Japanese, German, Spanish, 
and Arabic), the ordering
H $>$ S $>$ SEM persisted across all gain factors and in both
outcome measures (Fig.~\ref{fig:sensitivities}).
Mean values of $\gamma\%$ and ERSP$_\gamma$ vary approximately
linearly with $j$, yet the gaps between conditions remain nearly
constant.
At every $j$, the planned contrasts (Table~\ref{tab:sensitivity-range}) H $>$ S and S $>$ SEM remain
highly significant in each language under both perturbation schemes.
These results indicate that the core conclusion
\emph{Healthy(speech/semantics) $>$ SCZ(speech) $>$ SCZ(semantics)}
is structurally robust to reasonable variation in input gain and does
not rely on fine-tuned parameter choices across diverse linguistic contexts.

\begin{table*}[!htbp]
	\centering
	\scriptsize
	\caption{Sensitivity scan: paired-contrast design for perturbations of input gain $j$ across six TTS corpora ($n = 500$ per language). Entries indicate the range of $t$-statistics observed across the scan range $j \in \{0.8, \dots, 1.2\}$.}
	\label{tab:sensitivity-range}
	\resizebox{\textwidth}{!}{%
		\begin{tabular}{llccc}
			\toprule
			Language & Scheme & Contrast & $\gamma\%$ $t$-range & ERSP$_\gamma$ (dB) $t$-range \\
			\midrule
			\multirow{4}{*}{Mandarin}
			& Global (common-scale) & H $>$ S & $73.93$--$78.75$ & $37.29$--$57.57$ \\
			& Global (common-scale) & S $>$ SEM & $70.24$--$79.49$ & $54.93$--$58.22$ \\
			& Relative-scale (SCZ only) & H $>$ S & $61.08$--$94.48$ & $40.13$--$63.40$ \\
			& Relative-scale (SCZ only) & S $>$ SEM & $70.24$--$79.49$ & $54.93$--$58.22$ \\
			\midrule
			\multirow{4}{*}{English}
			& Global (common-scale) & H $>$ S & $37.60$--$71.48$ & $18.24$--$46.26$ \\
			& Global (common-scale) & S $>$ SEM & $60.69$--$73.22$ & $42.72$--$48.11$ \\
			& Relative-scale (SCZ only) & H $>$ S & $45.00$--$80.88$ & $24.53$--$45.92$ \\
			& Relative-scale (SCZ only) & S $>$ SEM & $60.69$--$73.22$ & $42.72$--$48.11$ \\
			\midrule
			\multirow{4}{*}{Japanese}
			& Global (common-scale) & H $>$ S & $86.36$--$95.16$ & $75.68$--$89.52$ \\
			& Global (common-scale) & S $>$ SEM & $67.16$--$95.08$ & $65.56$--$78.69$ \\
			& Relative-scale (SCZ only) & H $>$ S & $72.43$--$110.13$ & $68.53$--$92.15$ \\
			& Relative-scale (SCZ only) & S $>$ SEM & $67.16$--$95.08$ & $65.56$--$78.69$ \\
			\midrule
			\multirow{4}{*}{German}
			& Global (common-scale) & H $>$ S & $97.64$--$104.16$ & $80.59$--$112.51$ \\
			& Global (common-scale) & S $>$ SEM & $70.94$--$104.04$ & $75.77$--$85.24$ \\
			& Relative-scale (SCZ only) & H $>$ S & $80.33$--$120.68$ & $85.61$--$95.17$ \\
			& Relative-scale (SCZ only) & S $>$ SEM & $70.94$--$104.04$ & $75.77$--$85.24$ \\
			\midrule
			\multirow{4}{*}{Spanish}
			& Global (common-scale) & H $>$ S & $87.40$--$98.48$ & $77.63$--$99.08$ \\
			& Global (common-scale) & S $>$ SEM & $70.15$--$97.13$ & $65.58$--$84.61$ \\
			& Relative-scale (SCZ only) & H $>$ S & $74.63$--$111.23$ & $78.95$--$95.32$ \\
			& Relative-scale (SCZ only) & S $>$ SEM & $70.15$--$97.13$ & $65.58$--$84.61$ \\
			\midrule
			\multirow{4}{*}{Arabic}
			& Global (common-scale) & H $>$ S & $93.61$--$108.72$ & $79.11$--$126.47$ \\
			& Global (common-scale) & S $>$ SEM & $80.53$--$104.59$ & $72.58$--$95.64$ \\
			& Relative-scale (SCZ only) & H $>$ S & $84.33$--$123.66$ & $91.93$--$108.89$ \\
			& Relative-scale (SCZ only) & S $>$ SEM & $80.53$--$104.59$ & $72.58$--$95.64$ \\
			\bottomrule
		\end{tabular}
	}
\end{table*}

\subsection{Dynamical stability and bifurcation under speech drive}
\label{subsec:stability}

With all intrinsic model parameters held fixed, we performed a quasi-static stability scan for each utterance across all six languages by sweeping the externally constructed drive $I_{\mathrm{ext}}(t)$ over its empirical amplitude range. At every time point, we computed the maximal real part of the Jacobian eigenvalues to quantify the fraction of time the system spent in the positive (unstable) regime, denoted as \texttt{unstable\_frac}.

The stability analysis revealed a striking structural consistency across all linguistic corpora. The Healthy condition consistently maintained the neural dynamics in the unstable regime for a significantly longer duration than the SCZ--speech condition. Across the six languages, the mean \texttt{unstable\_frac} values for the Healthy group ranged from $0.493$ (Japanese) to $0.611$ (English), whereas the SCZ group ranged from $0.395$ (Japanese) to $0.495$ (English). Pairwise contrasts confirmed that these differences were highly significant in every language ($t > 104$, $p < 10^{-300}$, $n=500$ per language), with the Mandarin ($t=110.50$), English ($t=104.52$), Japanese ($t=108.16$), German ($t=110.44$), Spanish ($t=114.71$), and Arabic ($t=109.83$) corpora all exhibiting the same directional effect.

Furthermore, we examined the proximity of trajectories to bifurcation boundaries. No Hopf bifurcations were detected in any segment across the entire dataset. The saddle-node (SN) thresholds were tightly clustered and structurally invariant between groups: the median critical drive $I_{\mathrm{crit}}^{\mathrm{SN}}$ was approximately $0.11735$ for all conditions across all languages, with inter-quartile ranges fluctuating by less than $10^{-5}$.

These results indicate that, under identical macro-parameters and coupling strengths, the Healthy traces---driven by signals with higher RMS energy and richer spectral content---enter the nonlinear regime closer to the SN boundary more frequently. This dynamical behavior serves as the phase-plane counterpart to the spectral results observed in the $\gamma$-band. Since the intrinsic bifurcation manifolds (defined by $I_{\mathrm{crit}}^{\mathrm{SN}}$) are indistinguishable between groups, the robust reduction in $\gamma$-power and ERSP seen in SCZ conditions is attributable to the statistics of the external speech drive rather than a shift in the intrinsic instability of the E/I circuit.
\begin{figure*}[!htbp]
	\centering
	\begin{subfigure}[t]{0.48\textwidth}
		\centering
		\includegraphics[width=\linewidth]{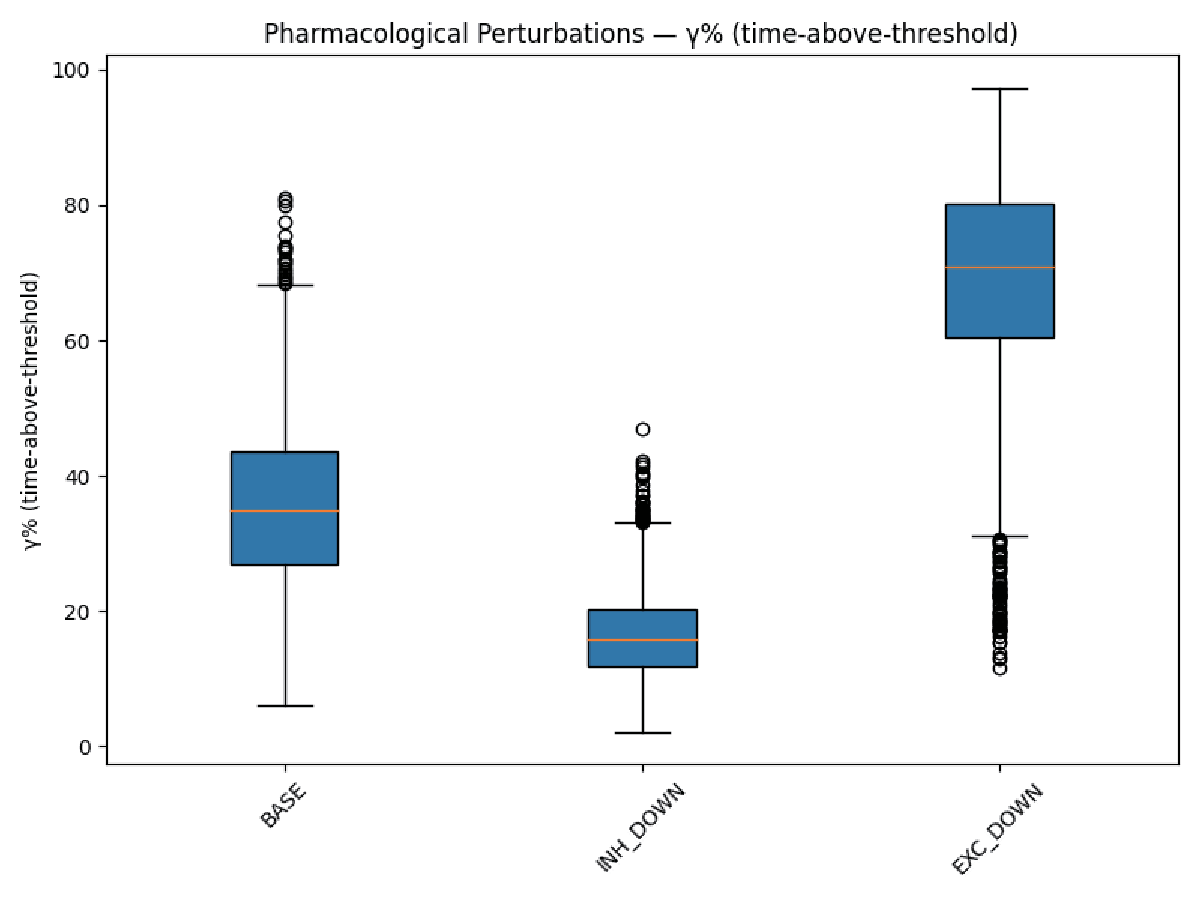}
		\caption{$\gamma\%$ (time-above-threshold) across all languages}
		\label{fig:pharma-gamma}
	\end{subfigure}\hfill
	\begin{subfigure}[t]{0.48\textwidth}
		\centering
		\includegraphics[width=\linewidth]{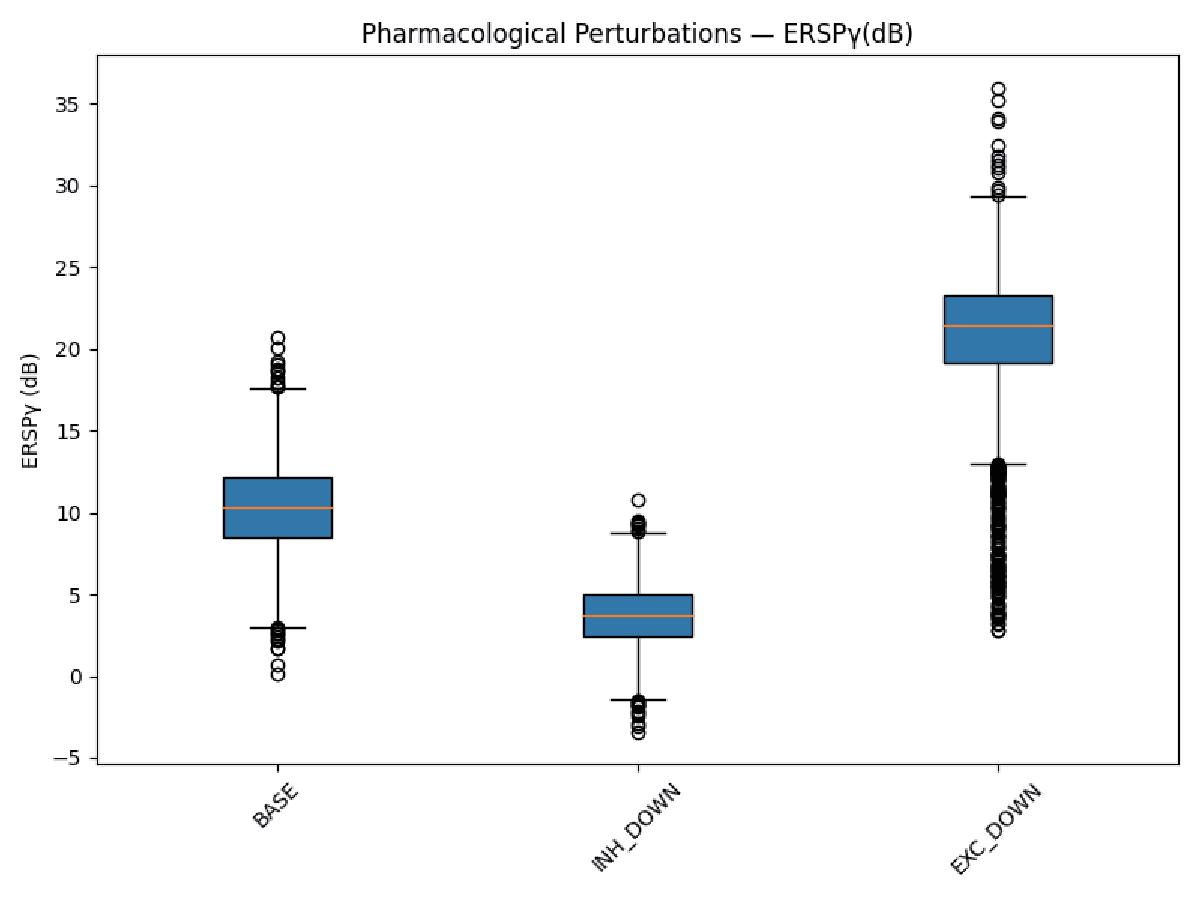}
		\caption{ERSP$_\gamma$ (dB) across all languages}
		\label{fig:pharma-ersp}
	\end{subfigure}
	
	\caption{Pharmacological perturbations across six languages.
		Each figure shows the combined results of all languages. 
		Left subpanel: $\gamma\%$ (time-above-threshold) for each condition. 
		Right subpanel: ERSP$_\gamma$ in dB for each condition. 
		Significant contrasts between Healthy $>$ SCZ and SCZ(speech) $>$ SCZ(semantics) across all languages are observed, demonstrating robustness in these effects across languages.}
	\label{fig:pharma-combined}
\end{figure*}

\begin{table*}[!htbp]
	\centering
	\scriptsize
	\caption{Pharmacological perturbations under equal-energy input for six TTS corpora ($n = 500$ segments per language). Paired contrasts vs.\ BASE. Entries report condition and baseline means; 95\% CIs are for the paired mean difference $\Delta$.}
	\label{tab:pharm}
	\resizebox{\textwidth}{!}{%
		\begin{tabular}{lllcccccc}
			\toprule
			Language & Contrast & Metric & Mean(Cond) & Mean(Base) & $\Delta$ & 95\% CI($\Delta$) & $t$ & $p$ (two-sided) \\
			\midrule
			\textbf{Mandarin} & INH\_DOWN $>$ BASE & $\gamma\%$ (\%) & 14.35 & 31.24 & $-16.89$ & [$-17.31$, $-16.47$] & $-79.18$ & $1.08\times10^{-284}$ \\
			& EXC\_DOWN $>$ BASE & $\gamma\%$ (\%) & 58.83 & 31.24 & $27.59$ & [$26.85$, $28.33$] & $73.23$ & $4.24\times10^{-269}$ \\
			& INH\_DOWN $>$ BASE & ERSP$_\gamma$ (dB) & 3.17 & 10.00 & $-6.83$ & [$-6.92$, $-6.75$] & $-157.35$ & $< 10^{-300}$ \\
			& EXC\_DOWN $>$ BASE & ERSP$_\gamma$ (dB) & 19.45 & 10.00 & $9.45$ & [$9.14$, $9.76$] & $59.45$ & $1.39\times10^{-228}$ \\
			\midrule
			\textbf{English} & INH\_DOWN $>$ BASE & $\gamma\%$ (\%) & 13.62 & 31.36 & $-17.74$ & [$-18.27$, $-17.22$] & $-66.13$ & $4.49\times10^{-249}$ \\
			& EXC\_DOWN $>$ BASE & $\gamma\%$ (\%) & 54.33 & 31.36 & $22.97$ & [$21.62$, $24.32$] & $33.46$ & $1.41\times10^{-129}$ \\
			& INH\_DOWN $>$ BASE & ERSP$_\gamma$ (dB) & 2.71 & 8.92 & $-6.21$ & [$-6.33$, $-6.09$] & $-103.08$ & $< 10^{-300}$ \\
			& EXC\_DOWN $>$ BASE & ERSP$_\gamma$ (dB) & 15.77 & 8.92 & $6.85$ & [$6.42$, $7.27$] & $31.51$ & $9.52\times10^{-121}$ \\
			\midrule
			\textbf{Japanese} & INH\_DOWN $>$ BASE & $\gamma\%$ (\%) & 15.08 & 33.41 & $-18.33$ & [$-18.92$, $-17.74$] & $-60.81$ & $6.90\times10^{-233}$ \\
			& EXC\_DOWN $>$ BASE & $\gamma\%$ (\%) & 69.74 & 33.41 & $36.33$ & [$35.31$, $37.34$] & $70.53$ & $1.16\times10^{-261}$ \\
			& INH\_DOWN $>$ BASE & ERSP$_\gamma$ (dB) & 3.39 & 10.16 & $-6.78$ & [$-6.88$, $-6.68$] & $-134.53$ & $< 10^{-300}$ \\
			& EXC\_DOWN $>$ BASE & ERSP$_\gamma$ (dB) & 21.69 & 10.16 & $11.53$ & [$11.20$, $11.85$] & $68.82$ & $7.48\times10^{-257}$ \\
			\midrule
			\textbf{German} & INH\_DOWN $>$ BASE & $\gamma\%$ (\%) & 20.31 & 41.93 & $-21.62$ & [$-22.21$, $-21.03$] & $-71.97$ & $1.16\times10^{-265}$ \\
			& EXC\_DOWN $>$ BASE & $\gamma\%$ (\%) & 78.33 & 41.93 & $36.40$ & [$35.35$, $37.45$] & $68.20$ & $4.45\times10^{-255}$ \\
			& INH\_DOWN $>$ BASE & ERSP$_\gamma$ (dB) & 4.69 & 11.36 & $-6.67$ & [$-6.77$, $-6.57$] & $-125.82$ & $< 10^{-300}$ \\
			& EXC\_DOWN $>$ BASE & ERSP$_\gamma$ (dB) & 22.98 & 11.36 & $11.62$ & [$11.27$, $11.97$] & $65.74$ & $6.48\times10^{-248}$ \\
			\midrule
			\textbf{Spanish} & INH\_DOWN $>$ BASE & $\gamma\%$ (\%) & 17.95 & 39.56 & $-21.62$ & [$-22.23$, $-21.00$] & $-69.00$ & $2.34\times10^{-257}$ \\
			& EXC\_DOWN $>$ BASE & $\gamma\%$ (\%) & 75.91 & 39.56 & $36.34$ & [$35.33$, $37.36$] & $70.18$ & $1.08\times10^{-260}$ \\
			& INH\_DOWN $>$ BASE & ERSP$_\gamma$ (dB) & 4.13 & 10.94 & $-6.82$ & [$-6.92$, $-6.72$] & $-136.06$ & $< 10^{-300}$ \\
			& EXC\_DOWN $>$ BASE & ERSP$_\gamma$ (dB) & 21.89 & 10.94 & $10.95$ & [$10.63$, $11.27$] & $67.31$ & $1.63\times10^{-252}$ \\
			\midrule
			\textbf{Arabic} & INH\_DOWN $>$ BASE & $\gamma\%$ (\%) & 17.47 & 37.27 & $-19.80$ & [$-20.37$, $-19.22$] & $-67.78$ & $7.07\times10^{-254}$ \\
			& EXC\_DOWN $>$ BASE & $\gamma\%$ (\%) & 76.53 & 37.27 & $39.26$ & [$38.22$, $40.29$] & $74.34$ & $4.42\times10^{-272}$ \\
			& INH\_DOWN $>$ BASE & ERSP$_\gamma$ (dB) & 3.98 & 10.48 & $-6.50$ & [$-6.60$, $-6.40$] & $-126.38$ & $< 10^{-300}$ \\
			& EXC\_DOWN $>$ BASE & ERSP$_\gamma$ (dB) & 23.03 & 10.48 & $12.56$ & [$12.20$, $12.91$] & $69.85$ & $9.33\times10^{-260}$ \\
			\bottomrule
		\end{tabular}%
	}
\end{table*}

\subsection{Pharmacological perturbations}\label{subsec:pharm}

We examined how the model responds to perturbations along the excitatory--inhibitory (E/I) axis across all six linguistic corpora.
We fixed the network architecture and noise seed, implementing two system-level manipulations that act in opposite directions on the effective coupling strengths:
\textit{INH\_DOWN} (reduced inhibition; mimicking GABA receptor antagonism by downscaling $\omega_{EI}$ and $\omega_{II}$) and
\textit{EXC\_DOWN} (reduced excitation; mimicking AMPA receptor antagonism by downscaling $\omega_{EE}$ and $\omega_{IE}$).

In all six languages (Mandarin, English, Japanese, German, Spanish, and Arabic), these manipulations produced clear, mutually corroborating directional effects (Fig.~\ref{fig:pharma-combined}).
Relative to baseline (BASE), the \textit{INH\_DOWN} condition markedly reduced $\gamma$ activity, significantly lowering both $\gamma\%$ and ERSP$_\gamma$.
Conversely, \textit{EXC\_DOWN} strongly amplified both metrics relative to BASE.
All planned paired contrasts (\textit{INH\_DOWN} vs.\ BASE and \textit{EXC\_DOWN} vs.\ BASE) were highly significant for both outcome measures across every language, with extremely large effect sizes (Table~\ref{tab:pharm}).
For instance, the reduction in ERSP$_\gamma$ under disinhibition (\textit{INH\_DOWN}) yielded $t$-statistics ranging from $-90$ (English) to over $-157$ (Mandarin), effectively reaching the limits of floating-point precision ($p < 10^{-300}$).
These bidirectional patterns accord with the mechanistic view that inhibitory circuits constrain and resonantly modulate $\gamma$ rhythms, and demonstrate that the model's sensitivity to E/I balance is a robust, language-invariant feature.

\section{Discussion}\label{sec:discussion}

Using a unified speech front end coupled to a Wilson--Cowan model, we systematically compared $\gamma$-band metrics across three conditions in matched multilingual TTS corpora spanning six languages (Mandarin, English, Japanese, German, Spanish, and Arabic) and arrive at three main conclusions. First, $\mathrm{ERSP}_\gamma$ and $\gamma\%$ exhibited a consistent hierarchy across all six languages (500 TTS segments per language)---Healthy (speech) $>$ SCZ (speech) $>$ SCZ (semantics)---with the two measures varying in the same direction and remaining stable across utterances. This pattern suggests that speech-evoked $\gamma$ differences primarily reflect shifts in the circuit’s E/I operating point rather than simple energy confounds \citep{Buzsaki2012,Pfurtscheller1999}. Operationally, these group and condition differences are implemented in
the model by a simple hierarchy of input gains applied to the same
speech-derived drive, with \(j_H = 1.00\), \(j_S = 0.75\), and
\(j_{\mathrm{SEM}} = 0.55\).
All structural parameters of the Wilson–Cowan network (E–I weights,
time constants, and noise amplitude) are shared across conditions, so the
only manipulation is the strength of the effective external drive
\(I(t)=j\,I_{\mathrm{ext}}(t)\).
We therefore interpret the gain parameter \(j\) as a phenomenological
measure of input sensitivity: it quantifies how strongly the same speech
envelope is transmitted through thalamo–cortical and cortico–cortical
pathways into the local E/I circuit.
Larger gains correspond to higher cortical sensitivity or salience
attribution to an identical physical input, whereas smaller gains
correspond to a relatively down-weighted, less effective drive.
The hierarchy \(j_H>j_S>j_{\mathrm{SEM}}\) encodes two assumptions that
are consistent with clinical findings: (i) relative to healthy listeners,
patients with schizophrenia have a reduced effective gain for speech,
contributing to attenuated 40-Hz ASSR power and \(\gamma\%\) at the group
level; and (ii) within the patient group, speech-like inputs still drive
the circuit more strongly than semantically scrambled controls, in line
with observations that natural speech remains perceptually salient even
when higher-level semantic integration is impaired.
This view links the scalar gains in the model to theories of abnormal
auditory gain control and salience attribution in schizophrenia, rather
than positing three qualitatively different network architectures for the
three conditions.
After equal-energy control (RMS pre-normalization), the ordering was fully preserved, supporting an explanation in terms of ``normalization--saturation,'' whereby effective gain together with the operating point governs the response \citep{CarandiniHeeger2012}. In gain-jitter experiments ($j\approx 0.8$--$1.2$), between-group gaps scaled approximately linearly while the ranking remained unchanged, indicating that the group differences are not driven by incidental noise or input-amplitude uncertainty but by structural shifts of the E/I operating point \citep{ErmentroutTerman2010}. The replication of the same hierarchy in typologically and prosodically diverse languages further suggests that these effects reflect language-general properties of the speech–circuit interaction rather than idiosyncrasies of a particular language. 

Under the row-sum constant and synchronous assumption, the network’s multi-channel LFP closely overlapped in time with the dynamics of an equivalent single node, with maximal discrepancies within numerical precision. This network--single-node equivalence substantially simplified subsequent stability and sensitivity analyses, allowing speech-driven scans to be recast as low-dimensional evaluations without loss of accuracy and lowering the technical barrier to linking $\gamma$ metrics directly to bifurcation structure \citep{ErmentroutTerman2010,Strogatz2015}. Stability and bifurcation analysis further showed that, at speech-drive intensities, trajectories are attracted to driven responses while operating more frequently in the nonlinear regime to the right of the saddle--node (SN) boundary, especially for the healthy speech, with no Hopf bifurcation detected across utterances. This interpretation is fully consistent with the measured time-above-threshold of local linear instability being larger for Healthy than for SCZ-speech.

Pharmacology-analog perturbations provided directional, causal evidence. Holding inputs and network topology fixed, reducing inhibitory weights ({INH\_DOWN}) significantly decreased $\gamma\%$ and $\mathrm{ERSP}_\gamma$, whereas reducing excitatory weights ({EXC\_DOWN}) significantly increased both metrics. This bidirectional control is consistent with the view that E/I balance determines whether $\gamma$ oscillations are generated or pushed away from resonance, and it aligns with the positive contribution of PV-mediated $\gamma$ rhythms to cortical circuit performance \citep{Sohal2009,Buzsaki2012,GonzalezBurgos2012}. Together with the stability and equivalence analyses, these results can be interpreted as predictable shifts of the operating point along the SN boundary (with no evidence of Hopf bifurcation), producing coherent upward or downward changes in $\gamma$ metrics \citep{ErmentroutTerman2010,Strogatz2015}. The fact that the same directional effects are obtained when applying the same perturbations to corpora in six different languages suggests that this operating-point picture is robust to cross-linguistic variation in phonotactics and prosody.

Several limitations remain. The study relies on a mean-field formulation with effective weights and does not explicitly model layer--column geometry, long-range interactions, or neurotransmitter kinetics, limiting our ability to address finer-grained questions such as cross-frequency coupling \citep{Wang2010,GiraudPoeppel2012}. In addition, although we used six languages, only naturalistic speech-like inputs were considered; future work could incorporate speech-in-noise, stimuli with controlled rhythm and phase, and non-speech control signals to more cleanly dissociate contributions of energy, structure, and temporal phase \citep{DingSimon2014,Hyafil2015}. Finally, aligning this framework with subject-level EEG/MEG---together with task manipulations or pharmacological interventions---will be important to further test the portability of the claim that $\gamma$ metrics primarily track the E/I operating point, and to explore potential applications in computational psychiatry \citep{Uhlhaas2010,GonzalezBurgos2012}.

\backmatter

\bmhead{Supplementary information}
Additional figures and simulation details are provided in the supplementary material.

\bmhead{Acknowledgements}
This work was supported by ...

\bmhead{Statements and Declarations}

\textbf{Funding} This work was supported by ...

\textbf{Competing interests} The authors declare that they have no competing interests.

\textbf{Data and code availability} Simulation code and processed data are available at ...

\textbf{Author contributions} FA designed the study and implemented the model; SA performed the simulations and analyses; both authors interpreted the results and wrote the manuscript.

\bibliographystyle{sn-mathphys-ay} 
\bibliography{refs}                

\end{document}